\RequirePackage{fix-cm}

\documentclass[smallextended]{svjour3}      

\smartqed  

\usepackage[font={small}]{caption}
\usepackage{dblfloatfix}  
\usepackage{subcaption}
\usepackage{mathtools}
\usepackage{multirow}
\usepackage{graphicx}  
\usepackage{enumitem}
\usepackage{relsize}
\usepackage{amsmath}
\usepackage{amssymb}
\usepackage{url}

\usepackage{xcolor}
\definecolor{boxgrey}{HTML}{F3F3F3}

\newcommand{\hlbox}[2]{%
  \begin{center}%
    \fcolorbox{white}{boxgrey}{%
      \parbox{.9\columnwidth}{\noindent \textbf{#1}. \textit{#2}}
    }%
  \end{center}%
}

\begin{document}

        \vspace*{1cm}
            
        \noindent \textbf{Interplay between Upsampling and Regularization for Provider Fairness in Recommender Systems}
            
        \vspace{2cm}
        
        \noindent Journal Item
            
        \vspace{2cm}
 
        \hrule
        
        \vspace{0.5cm}
        
        \noindent \textbf{How to cite:}
        
        \vspace{2mm} 
        
        \noindent Boratto, L., Fenu, G., \& Marras, M. (2021). Interplay between Upsampling and Regularization for Provider Fairness in Recommender Systems. User Model User-Adap Inter. \url{https://doi.org/10.1007/s11257-021-09294-8}
        
        \vspace{0.5cm}
        
        \hrule
        
        \vspace{0.5cm}
            
        \noindent Version: Accepted Manuscript

\title{Interplay between Upsampling and Regularization for Provider Fairness in Recommender Systems}
\titlerunning{Interplay between Upsampling and Regularization for Provider Fairness}
\journalname{User Model User-Adap Inter.}

\author{Ludovico Boratto \and Gianni Fenu \and \\Mirko Marras }

\institute{
           Ludovico Boratto \at
           EURECAT, Centre Tecnol\'ogic de Catalunya, Barcelona, Spain \\
           \email{ludovico.boratto@acm.org}    
           \and 
           Gianni Fenu \at
           University of Cagliari, Cagliari, Italy \\
           \email{fenu@unica.it}       
           \and 
           Mirko Marras * \at
           \'Ecole Polytechnique F\'ed\'erale de Lausanne (EPFL), Lausanne, Switzerland \\
           \email{mirko.marras@acm.org}    \\
           * Work conducted while the author was at University of Cagliari, Cagliari, Italy
}

\date{Received: date / Accepted: date}

\maketitle

\begin{abstract}
Considering the impact of recommendations on item providers is one of the duties of multi-sided recommender systems. Item providers are key stakeholders in online platforms, and their earnings and plans are influenced by the exposure their items receive in recommended lists. Prior work showed that certain minority groups of providers, characterized by a common sensitive attribute (e.g., gender or race), are being disproportionately affected by indirect and unintentional discrimination. Our study in this paper handles a situation where ($i$) the same provider is associated with multiple items of a list suggested to a user, ($ii$) an item is created by more than one provider jointly, and ($iii$) predicted user-item relevance scores are biasedly estimated for items of provider groups. Under this scenario, we assess disparities in relevance, visibility, and exposure, by simulating diverse representations of the minority group in the catalog and the interactions. Based on emerged unfair outcomes, we devise a treatment that combines observation upsampling and loss regularization, while learning user-item relevance scores. Experiments on real-world data demonstrate that our treatment leads to lower disparate relevance. The resulting recommended lists show fairer visibility and exposure, higher minority item coverage, and negligible loss in recommendation utility.  
\keywords{Recommender Systems \and Collaborative Filtering \and Multi-Stakeholder \and Fairness \and Exposure \and Visibility \and Regularization}
\end{abstract}

\section{Introduction} \label{sec:introduction}
Recommender systems help individuals explore vast catalogs of items. To this end, such systems adopt a model that implements a suitable way of ranking items. Conventionally, items are ranked in order of their decreasing relevance for a given user, \emph{estimated via machine learning}. The literature traditionally focused on optimizing user-item relevance for user's recommendation utility~\cite{DBLP:reference/sp/RicciRS15}. However, many recommendation scenarios involve multiple stakeholders, and should account for the impact on more than one group of participants~\cite{DBLP:journals/corr/Burke17aa,DBLP:journals/ipm/BorattoFM21}. For instance, the ranked lists may influence profits and plans of item providers~\cite{DBLP:journals/tmis/JannachJ19}. 

\vspace{2mm} \noindent \textbf{Context}. The motivation driving this paper is that a model, optimized for user's recommendation utility, can introduce an indirect and unintentional \emph{discrimination} for providers belonging to a \emph{legally-protected minority class} (e.g., when considering gender or ethnicity as a sensitive attribute)~\cite{DBLP:journals/datamine/Zliobaite17,DBLP:conf/innovations/DworkHPRZ12}. Given the primary role of recommender systems also for minority providers, having their items unfairly recommended would have human, ethical, social, and economic consequences~\cite{DBLP:reference/sp/RicciRS15}. Furthermore, due to these phenomena, providers might lose their trust in the platform and consequently leave it, impacting on the ecosystem as a whole. Hence, it is imperative to uncover, characterize, and mitigate discrimination inherent in the recommendation model, so that no platform systematically and repeatedly disadvantages minority providers.

\vspace{2mm} \noindent \textbf{Problem Statement}. The literature in ranking and recommendation recently focused on aligning the exposure or the attention to providers with their relevance or contribution in the catalog, at individual or group level~\cite{DBLP:conf/ssdbm/YangS17,DBLP:journals/ajiips/LiuSZGX19,DBLP:conf/fat/KamishimaAAS18,DBLP:conf/sigir/BiegaGW18}. Our study encodes the idea of a group-level proportionality between the contribution in the catalog and the relevance, the visibility, and the exposure, following a distributive norm based on \emph{equity}~\cite{DBLP:book/Walster1973new}. Operationalizing this notion during the user-item relevance optimization stage may be envisioned as a \emph{proactive way} of addressing provider's fairness along the recommendation pipeline, given that relevance scores are the input for the final ranking stage. Being optimized for their ability to rank, the estimated relevance scores directly influence the chance of an item being ranked high (i.e., the higher the relevance is, the more likely the item appears at the top). If these relevances are biased against the minority group, the recommender system is unfairly giving minority items less chance of being ranked high. Given its connection with the final ranking, relevance is thus an \emph{internal algorithmic asset} to be allocated to provider groups, and not just a property of user-item pairs to be estimated. Therefore, controlling relevances of items of a provider group can be a driver of recommendation outcomes with lower disparities. 

Despite potentially bringing fairness-related benefits on the suggested lists by itself, controlling predicted relevance scores may also help to deal with situations where true expected relevances required by existing fairness-aware treatments are not available. Ensuring that a model improves its capability to deem the items of the minority as relevant is not trivial, since minority items tend to be under-represented in interactions. This may influence the predicted relevance and, in cascade, the recommendations involving minority providers. The disparate impact we address consists in items of \emph{a small minority group of providers} systematically receiving unfairly low relevance and, by extension, an exposure not proportional to their contribution in the catalog. Our goal in this paper is thus to investigate whether, during the learning stage, taking actions for increasing the relevance of items from the minority group positively impacts on providers' group fairness in recommendations.  

\vspace{2mm} \noindent \textbf{Open Issues}. While a range of frameworks to assess and mitigate provider unfairness has been introduced in the context of non-personalized people rankings~\cite{DBLP:conf/sigir/BiegaGW18,DBLP:conf/kdd/SinghJ18,DBLP:journals/pvldb/LahotiGW19} and item recommendation~\cite{DBLP:conf/fat/KamishimaAAS18,DBLP:conf/kdd/BeutelCDQWWHZHC19}, several issues remain open. 

Despite being extendable to many-to-many item-provider associations, existing frameworks for provider fairness have been assessed on settings with a one-to-one association between items and providers~\cite{DBLP:conf/kdd/BeutelCDQWWHZHC19,DBLP:journals/corr/abs-1901-10437}. This is natural in a people ranking, since the concepts of provider and item being ranked coincide. However, under a more general item recommendation scenario, items and providers may be linked by a many-to-many relationship (e.g., a movie having multiple directors or a director offering multiple movies). Hence, there is a need to assess how fair are recommendations for providers in the general context we described (e.g., for items having both female and male providers). 

Furthermore, disparate exposure has been traditionally mitigated through a form of re-ranking, assuming to have access to true unbiased relevances~\cite{DBLP:conf/kdd/SinghJ18,DBLP:conf/sigir/BiegaGW18}. However, these relevances are typically estimated by means of a machine-learning technique, leading to a possibly biased value of the relevance scores. Indeed, recommender systems are known to be biased from several perspectives (e.g., popularity, presentation, and, obviously, unfairness for users and providers). Predicting a relevance score on biased/unfair results and basing a re-ranking approach on a possibly biased relevance may lead to undesired effects, considering that relevance directly influences the chance of an item being ranked high. This issue is urging novel methods able to instill a fine-grained share of relevance across groups in the algorithmic mechanics. This would generate a \emph{tangible impact on disparity reduction in the final ranking}.  

To the best of our knowledge, no approach deals with controlling the balance of relevance estimations across provider groups, under the above scenario. Indeed, while in-processing regularizations of relevance exist~\cite{DBLP:conf/fat/KamishimaAAS18,DBLP:conf/kdd/BeutelCDQWWHZHC19} and this would overcome the second issue, these treatments are fundamentally driven by a fairness objective different from ours, not relying on controlling the share of relevances, and still assessed on a one-to-one item-provider relationship. 

\vspace{2mm} \noindent \textbf{Motivating Intuitions}. The intuitions that drive our approach are depicted with concrete examples, taken from the MovieLens-10M dataset, presented in detail in Section~\ref{sec:dataset}. Considering a binary gender attribute\footnote{While gender is by no means a binary construct, to the best of our knowledge no dataset with non-binary genders exists. What we are considering is a binary feature, as the current publicly available datasets offer.} and using movie directors as a proxy of providers, female directors appear on the $6.0\%$ of items in the catalog, but end up being under-represented with only $3.9\%$ of interactions. With the pair-wise approach we employed in this work and the (un)fairness metric we will present, female providers receive $2.9\%$ of the total item relevance (and $2.8\%$ of exposure), being affected by the disparate impact. 

Considering that the items having more interactions  are more likely to receive high relevance and be recommended at the top of the ranking (i.e., the well-known popularity phenomenon), we investigated whether upsampling the interactions involving the minority group of providers, to reach a percentage aligned with their representation in the catalog (i.e., $6.0\%$ of the total interactions), can reduce disparities in relevance and, by extension, in exposure. Giving the upsampled set of interactions as an input to the same pair-wise algorithm led to female providers receiving $5.4\%$ of relevance and $5.2\%$ of exposure, still far from the $6.0\%$ of representation in the catalog. Given this gap, we then regularized for the share of relevances during the learning process, leveraging upsampled interactions (which are important to enable the regularization). This latter setting led to a $5.9\%$ of relevance and $5.8\%$ of exposure for female providers, reducing the initial disparate impact. These preliminary practical results motivated us to investigate how upsampling and regularization can lead to higher relevance and lower disparate exposure for the minority. 

\vspace{2mm} \noindent \textbf{Contributions}. Compared to prior work, both in the fairness metric and the mitigation, we consider a many-to-many relationship between items and their providers, and assess the representation of each value of a sensitive attribute in a given item (i.e., we would assess how represented each gender is in that item). Under this scenario, to reduce disparities in relevance and exposure, we propose a pre-processing strategy that up-samples interactions where the minority group is predominant (e.g., an item where the minority is represented with two providers is better than item with only one provider of that group; moreover, the lower the representation of the majority in that item is, the more we can help the minority, by favoring an upsampling of these latter items). In addition, an in-processing component aims to control that the relevance given to the items of the minority group is proportional to the minority group contribution in the catalog. 
Our contribution is summarized as follows:

\begin{itemize}
	\item we characterize disparities in predicted relevance, visibility, and exposure against the minority group of providers, and assess their existence on synthetic data that simulates diverse representations of the group in the catalog and the interactions, learning lessons that guide our mitigation;
	\item we present a mitigation approach that relies on (i) tailored upsampling in pre-processing and (ii) a regularization term added to the original training optimization function, to operationalize our motivating intuitions;
	\item we leverage two public datasets with gender information of the providers, enabling the consequent evaluation of the impact of our metrics and strategies on real-world datasets with very small minority groups.
\end{itemize}

\noindent \textbf{Roadmap}. The remaining of this paper is structured as follows: Section~\ref{sec:concepts-definitions} formalizes key concepts and metrics, and Section~\ref{sec:exploration} describes our exploratory analysis. Then, Section~\ref{sec:mitigation} introduces our mitigation approach, while Section~\ref{sec:experimental-evaluation} assess its feasibility. Section~\ref{sec:related-work} provides connections with prior work. Finally,  Section~\ref{sec:conclusions} depicts concluding remarks and future research directions.

\section{Concepts and Definitions} \label{sec:concepts-definitions}
In this section, we outline the recommendation scenario we seek to investigate and the concepts and definitions used throughout this paper.

\subsection{Recommender System Formalization}\label{sec:recsys-formalization}
Given a set of users $U$, a set of items $I$, and a set of providers $P$, we assume that each item $i \in I$ is jointly offered by a subset of providers $P_i \subset P$, with $|P_i| > 0$, and a provider $p \in P$ offers a subset of items $I_p \subset I$, with $|I_p| > 0$. For instance, in the context of course recommendation, if we consider instructors as providers of course items, a course could have two instructors who give lectures cooperatively. Similarly, the same instructor could deliver three different courses on the platform, two of them cooperatively and one alone, just as an example. Each provider $p \in P$ is associated with $N$ discrete sensitive attributes $(a_1^p, a_2^p, \cdots, a_n^p)$, with $a_1^p \in A_1 \subset  \mathbb{N}$, $\ldots$, $a_n^p \in A_n \subset  \mathbb{N}$. For instance, a set $A_{j}$ could be associated with the gender attribute and, thus, being defined as $A_j = \{0:female, 1:male, \dots\}$, assuming that an attribute is discrete and that we encoded each discrete value to a unique integer. 

We assume that users have interacted with a subset of items in $I$. The collected feedback from user-item interactions can be abstracted to a set of pairs ($u$, $i$) obtained from the normal user's activity or triplets ($u$, $i$, $value$), whose $value$ is either provided by users (e.g., ratings) or computed by the system (e.g, frequency). In our study, we consider pairs derived from explicit feedback, by applying a pre-selected threshold to rating values, in order to model the recommendation task as a personalized ranking problem. We denote the user-item feedback matrix by $R \in \mathbb{R}^{|U|*|I|}$, where $R_{u,i} > 0$  indicates that user $u$ interacted with item $i$, and $R_{u,i}=0$ otherwise. Furthermore, we denote the set of items that user $u\in U$ interacted with by $I_u=\{i\in I\,:\,R_{u,i} > 0\}$. 

We assume that each user $u \in U$ and item $i \in I$ is internally represented by a $D$-sized numerical vector from a user-vector matrix $W$ and an item-vector matrix $X$, respectively. The recommender system's task is to optimize $\theta = (W,X)$ for predicting unobserved user-item relevance. It can be abstracted as learning $\widetilde{R}_{u,i} = f_{\theta}(u,i)$, where $\widetilde{R}_{u,i}$ denotes the predicted relevance, $\theta$ denotes learnt user and item matrices, and $f$ denotes the function predicting the relevance between $W_u$ and $X_i$. Given a user $u$, items $i \in I \setminus I_u$ are ranked by decreasing $\widetilde{R}_{u,i}$, and top-$k$, with $k\in\mathbb N$ and $k>0$, items are recommended. Our study will focus on $k=10$ recommendations per user, since they probably get the most attention of users and 10 is a widely employed cut-off. Finally, we denote the set of $k\in\mathbb N$ items recommended to user $u$ by $\Tilde{I}_u$. 

\subsection{Associating Providers' Sensitive Attributes to Items} \label{sec:sens-attr-items}
Formalizing our target notion of fairness for provider groups, under the scenario depicted in Section \ref{sec:recsys-formalization}, requires to deal with several aspects. Fairness studies in ranking and recommendation traditionally targeted people as entities to be ranked or recommended~\cite{DBLP:conf/sigir/BiegaGW18,DBLP:conf/ssdbm/YangS17,DBLP:journals/pvldb/LahotiGW19}. While still having individuals being directly affected by how recommendations are generated, entities to be recommended are not always individuals, and may include items (e.g., movies, courses). This turns out to key challenges that rise in cascade.

First, in many cases, there is no direct one-to-one mapping between an item and the individual who has created or offered it (i.e., the provider). Realistic scenarios need to consider items created by more than one provider cooperatively (e.g., a course with two instructors) and how the sensitive attributes are associated to the involved providers. It can be even difficult to come up with a one-to-many mapping for items offered by an entity not directly linked to individuals (e.g., a training company providing an online course).

Second, the fact that an item might have more than one provider behind it poses the problem of how to model the representation of a providers' sensitive attribute, when considering that item (e.g., how each gender is represented in a given item), based on the individuals associated to it. Linking a unique variable, either binary or multi-class, discrete or continuous, to a sensitive attribute of a provider and claim fairness on such a variable is often impractical. More sophisticated solutions should be considered. For instance, the metrics proposed by Biega et al. in \cite{DBLP:conf/sigir/BiegaGW18} and in the TREC Fair Track \cite{DBLP:journals/corr/abs-2003-11650} have been devised to handle items with multiple providers with different attributes.

Based on these observations, we define a notion of \texttt{sensitive attribute representation} for an item $i$, subjected to a sensitive attribute $A$. This notion requires to consider the membership of each provider $p \in P_i$ to a class of the sensitive attribute $A$ (which we previously denoted as $a^p$), while mapping sensitive attributes to items.

\begin{definition}[Sensitive attribute representation]
  Given a sensitive attribute $A \subset \mathbb{N}$, the sensitive attribute representation $s_i^A$ of an item $i$ with respect to $A$ is defined as: 
  \begin{equation}
      s_i^A = [ \; |P_i^a| \;, \; \forall \; a \in A]
  \end{equation}
  \label{eq:sensitivity}
\end{definition}  

where $P_i^a$ is the set of $i$'s providers with attribute $a \in A$. Each vector $s_i^A$ has size $|A|$ for all items $i \in I$, and each of its values represents the number of providers who belong to a given class of the attribute $A$, ranging in $[0, |P_i|]$. Similarly to us, Sapyezinski et al.~\cite{DBLP:conf/ssdbm/YangS17} use a function to map each ranked item to a vector. However, their vector is used as a proxy of uncertainty, while assigning a sensitive attribute value to a person to be ranked (e.g., given a binary gender construct, if a system considers that a person is male with a probability of $10\%$, the vector associated to that person is $[0.10,0.90]$). Our notion differs both conceptually and operationally, as we model and compute how each value a sensitive attribute can assume is represented across providers associated to a given item, in magnitude. Furthermore, our notion could be extended to model uncertainty, while getting the value of the sensitive attribute associated with a provider, assumed by us to be $a \in A \subset \mathbb{N}$. To better highlight our contribution, our study leaves this combination as a future work.   

\subsection{Identifying the Minority Group}
Our study considers groups of providers who belong to a given class of the attribute $a \in A$. Each group is involved in the creation/delivering of a certain number of items in the catalog and, consequently, in a certain number of the user-item interactions. Specifically, given the definitions previously  provided in Section \ref{sec:sens-attr-items}, the representation of a group in the catalog and the interactions is computed in our study as follows:  

\begin{definition}[Provider group representation in the catalog]
  Given a sensitive attribute $A \subset \mathbb{N}$, the representation of providers with a value of the sensitive attribute $a \in A$ in the \emph{catalog}, is defined as: 
  \begin{equation}
      \mathcal{C}^a = \frac{1}{|I|} \sum_{i \in I} \frac{s_i^A(a)}{\sum_{\widetilde{a} \in A} s_i^A(\widetilde{a})} 
  \end{equation}
  \label{eq:contribution}
\end{definition} 

where $s_i^A(a)$ is the element of the vector $s_i^A$ associated to the value $a$, as per definition in Eq. \ref{eq:sensitivity}. The representation $\mathcal{C}^a$ ranges in $[0,1]$, and accounts for the contribution of providers belonging to a given group in the delivering of items in the catalog. A value close to $0$ means that $a$'s providers rarely contribute to items in the catalog, and viceversa for values close to $1$. Similarly, we define the representation of a provider group in the interactions.

\begin{definition}[Provider group representation in the interactions]
  Given a sensitive attribute $A \subset \mathbb{N}$, the representation of providers with a value of the sensitive attribute equal to $a \in A$, in the \emph{interactions} $R$, where $M = \{(u,i) \; : \; R_{u,i}>0\}$ are the observed interactions, is defined as: 
  \begin{equation}
      \mathcal{O}^a = \frac{1}{|M|} \sum_{(u,i) \in M} \frac{s_i^A(a)}{\sum_{\widetilde{a} \in A} s_i^A(\widetilde{a})} 
  \end{equation}
\end{definition} 

In our study, we are interested in investigating how recommendation decisions impact on a group of providers identified as a minority. There exists different modalities to identify a minority group $\text{a}_{\min}$, one of them being \emph{the lowest representation in the catalog}, i.e., $\text{a}_{\min} = argmin_{a \in A} \mathcal{C}_a$. This choice will better support us to account for differences in contribution among provider groups, assuming that the catalog curation does not suffer from sampling bias (e.g., a course platform that refuses to add courses given by female teachers to its catalog). While it could be reasonable to assume that certain groups of providers are less represented than others in the catalog (e.g., because certain categories of items are traditionally offered by providers of a given gender), the recommendation loop may lead to under-represent the minority group in the interactions more and more with respect to its group contribution in the catalog, i.e., $\mathcal{C}^{\text{a}_{\min}} > \mathcal{O}^{\text{a}_{\min}}$. This effect may inadvertently bias the learnt relevance, and, consequently, detain recommendations of minority-group items. 

\subsection{Formalizing Disparities}\label{sec:disp-impacts}
To assess the extent to which the recommender system generates disparities, we define three core disparity metrics. One of them considers an internal perspective and monitors the difference of predicted relevance between providers' groups. The other metrics operate on the final outcomes of the recommender system, monitoring differences in visibility and exposure. 

More precisely, the disparity in relevance ($\Delta \mathcal{R}$) is quantified as the absolute difference between the representation in the catalog ($\mathcal{C}^{\text{a}_{\min}}$) and the percentage of relevance for the minority group:  

\begin{equation}\label{eq:disp-rel}
    \Delta \mathcal{R} = \left|\frac{1}{|U|} \sum_{u \in U} \frac{\sum_{\text{pos}=1}^{k} \widetilde{R}_{u,\rho_{\theta}(u,\text{pos})} \cdot s_{\rho_{\theta}(u,\text{pos})}^A({\text{a}_{\min}})}{\sum_{\text{pos}=1}^{k} \sum_{a \in A} \widetilde{R}_{u,\rho_{\theta}(u,\text{pos})} \cdot  s_{\rho_{\theta}(u,\text{pos})}^A(a)} - \mathcal{C}^{\text{a}_{\min}}\right|
\end{equation}
\vspace{1mm} 

\vspace{2mm} \noindent where $\rho_{\theta}(u,\text{pos})$ represents the item recommended at position $\text{pos}$ for users $u$ and  $\widetilde{R}_{u,\rho_{\theta}(u,\text{pos})}$ refers to the predicted relevance formalized in Section \ref{sec:recsys-formalization}, while the terms $\mathcal{C}^{\text{a}_{\min}}$ and $s_{\rho_{\theta}(u,p)}^A$ derive from Eqs. \ref{eq:contribution} and \ref{eq:sensitivity}, respectively. Scores of $\Delta \mathcal{R}$ refer to top-$k$ recommendations and range in $[0,1]$, with higher values indicating a higher disparity of the degree of relevance estimates with respect to the contribution in the catalog for the minority group.

A disparity in relevances might not necessarily imply that the minority group is discriminated based on its exposure or visibility in the recommendations lists \cite{DBLP:conf/kdd/SinghJ18}, which is exactly what we aim to investigate in this paper. For this reason, we also define the difference between the contribution in the catalog and the percentage of visibility ($\Delta \mathcal{V}$) and of exposure ($\Delta \mathcal{E}$) for items of the minority group. Disparate visibility and exposure are formalized as follows:
 
\begin{equation}
    \Delta \mathcal{V} = \left|\frac{1}{|U|} \sum_{u \in U} \frac{\sum_{\text{pos}=1}^{k} s_{\rho_{\theta}(u,\text{pos})}^A({\text{a}_{\min}})}{\sum_{\text{pos}=1}^{k} \sum_{a \in A} s_{\rho_{\theta}(u,\text{pos})}^A(a)} - \mathcal{C}^{\text{a}_{\min}}\right|
    \label{eq:dv}
\end{equation}
\vspace{1mm} 
\begin{equation}
    \Delta \mathcal{E} = \left|\frac{1}{|U|} \sum_{u \in U} \frac{\sum_{\text{pos}=1}^{k} \frac{1}{log_2(\text{pos}+1)} s_{\rho_{\theta}(u,\text{pos})}^A({\text{a}_{\min}})}{\sum_{\text{pos}=1}^{k} \sum_{a \in A}  \frac{1}{log_2(\text{pos})} s_{\rho_{\theta}(u,\text{pos})}^A(a)} - \mathcal{C}^{\text{a}_{\min}} \right|
    \label{eq:de}
\end{equation}

\vspace{2mm} \noindent where, $\rho$, $\widetilde{R}$, $\mathcal{C}^{\text{a}_{\min}}$, and $s_{\rho_{\theta}(u,\text{pos})}^A$ are defined as above. Scores of $\Delta \mathcal{V}$ and $\Delta \mathcal{E}$ refer to top-$k$ recommendations and range in $[0,1]$, with lower values indicating a lower disparity w.r.t. the contribution in the catalog.

\section{Optimizing under Different Catalog-Interaction Representations} \label{sec:exploration}
To illustrate the unfairness against a minority group of providers and further emphasize the value of our analytical modeling, we simulate various imbalances in catalog and interactions, for the minority group. Then, we characterize to what extent a model is unfair against the minority group. Specifically, the exploratory study presented in this section aims to assess the extent to which the share of relevance across groups depends on \emph{imbalances} between catalog and observation representations, and whether reducing the degree of imbalance between the representations in the catalog and the interactions for a minority group leads to lower disparate exposure. Our hypothesis is that there is a strongly direct relationship between the imbalance in catalog-observation representations and the estimated disparities defined in Section \ref{sec:disp-impacts}. 

\subsection{Pair-wise Optimization and Exploratory Protocols}
Pair-wise optimization is one of the most influential approaches to train recommendation models and represents the foundation of many cutting edge personalized algorithms~\cite{chen2017attentive,xue2017deep,xiao2017attentional}. The underlying Bayesian formulation~\cite{rendle2012bpr} aims to maximize a posterior probability that can be adapted to the parameter vector of an arbitrary model class (e.g., matrix factorization or neighborhood-based). In our study, we adopt matrix factorization~\cite{koren2009matrix}, due to its popularity and flexibility. Model parameters $\theta$, i.e., user and item matrices, are estimated through an objective function that maximizes the margin between (i) the relevance $f_{\theta}(u,i)$ predicted for an observed item $i$ and, (ii) the relevance $f_{\theta}(u,j)$ predicted for an unobserved item $j$. The optimization process considers a set of triplets $D$ that are fed into the model during training: 

\vspace{-2mm}

\begin{equation}
    D = \{(u,i,j) \, | \, u \in U, i \in I^{+}_{u}, j \in I^{-}_{u}\}
\end{equation}

\vspace{2mm}

\noindent where $I^{+}_{u}$ and $I^{-}_{u}$ are the sets of items for which user $u$'s feedback is observed and unobserved, respectively. 

The original implementation proposed by \cite{rendle2012bpr} requires that, for each user $u$, triplets $(u,i,j)$ per observed item $i$ should be created; the unobserved item $j$ is randomly selected. The objective function can be formalized as follows:  

\begin{equation}
\underset{\theta}{\operatorname{argmax}} \mathop{\sum}_{(u,i,j) \in D} \delta{(f_{\theta}(u,i) - f_{\theta}(u,j)) - \left\lVert \theta \right\rVert_2^2} 
\label{eq:pair-wise}
\end{equation}

\vspace{2mm}

\noindent where $\delta$ is a sigmoid function returning a value between 0 and 1.

The code for our study was implemented in \emph{Python} on top of \emph{Tensorflow}. User and item matrices, with vectors of size $100$, were initialized with values uniformly distributed in $[0,1]$. The optimization function is transformed to the equivalent minimization dual problem. For each user, we randomly took apart $70\%$ of their interactions for training, $10\%$ for validation, and $20\%$ for testing. Given the training user-item interactions, the model was served with batches of $1,024$ triplets. For each user $u$, we created $10$ triplets $(u,i,j)$ per observed item $i$; the unobserved item $j$ was randomly selected for each triplet. The optimizer used for gradient update was \textit{Adam}. Training lasted until convergence on the validation set. Parameters were selected via grid search on the validation set. 

\subsection{Observations on Synthetic Datasets}
To investigate if and to what extent the share of relevance across providers' groups depends on \emph{imbalances} between catalog and observation representations, we consider a recommendation context that associates each provider $p \in P$ with a generic binary sensitive attribute, and we assume that each item is associated with a single provider, leaving experiments on items associated with more than one provider to the real-world datasets leveraged in Section \ref{sec:experimental-evaluation}. 

Specifically, the imbalances considered in this study are subdivided in two forms: \texttt{catalog imbalance} and \texttt{observation imbalance}. Catalog imbalances emerge when providers from a different group occur in the catalog with varied frequencies. For instance, there may be significantly fewer female/male providers than male/female providers who offer items to users. On the other hand, with observation imbalances, users may interact with items from certain provider groups with different tendencies. This imbalance is often part of a feedback loop involving existing methods of recommendation, either introduced by models or by humans. If users do not receive any item offered by a provider belonging to a certain group, users will not interact with that class of providers. In cascade, models will be served with only few data on this preference relation. For instance, train data about female/male providers may be significantly less than train data about male/female providers. 

\begin{figure}[!t]
\begin{subfigure}[t]{0.48\linewidth}
    \centering
    \includegraphics[width=1.0\linewidth]{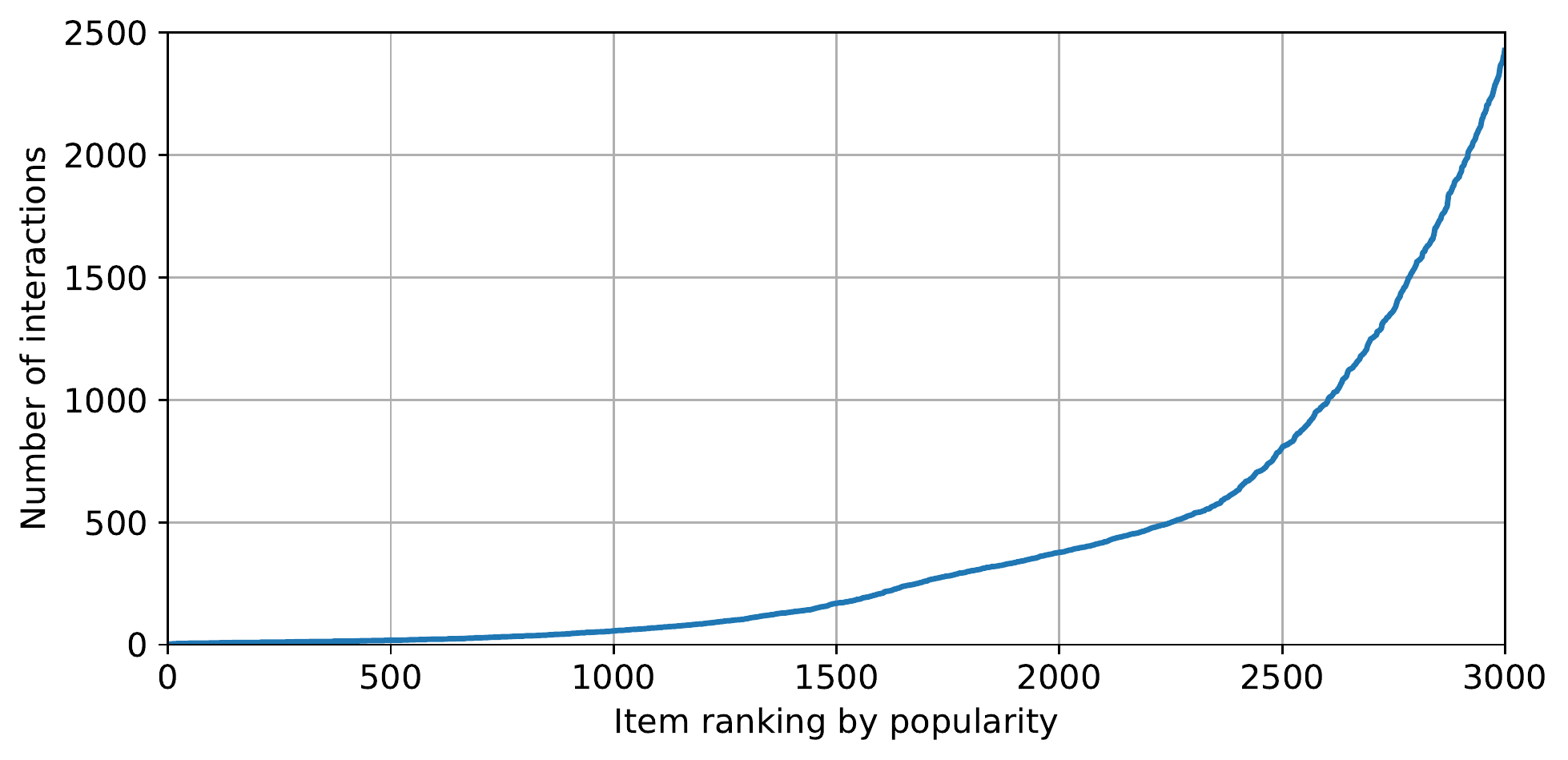}
    \caption{Popularity tail}
\end{subfigure}
\begin{subfigure}[t]{0.48\linewidth}
    \centering
    \includegraphics[width=1.0\linewidth]{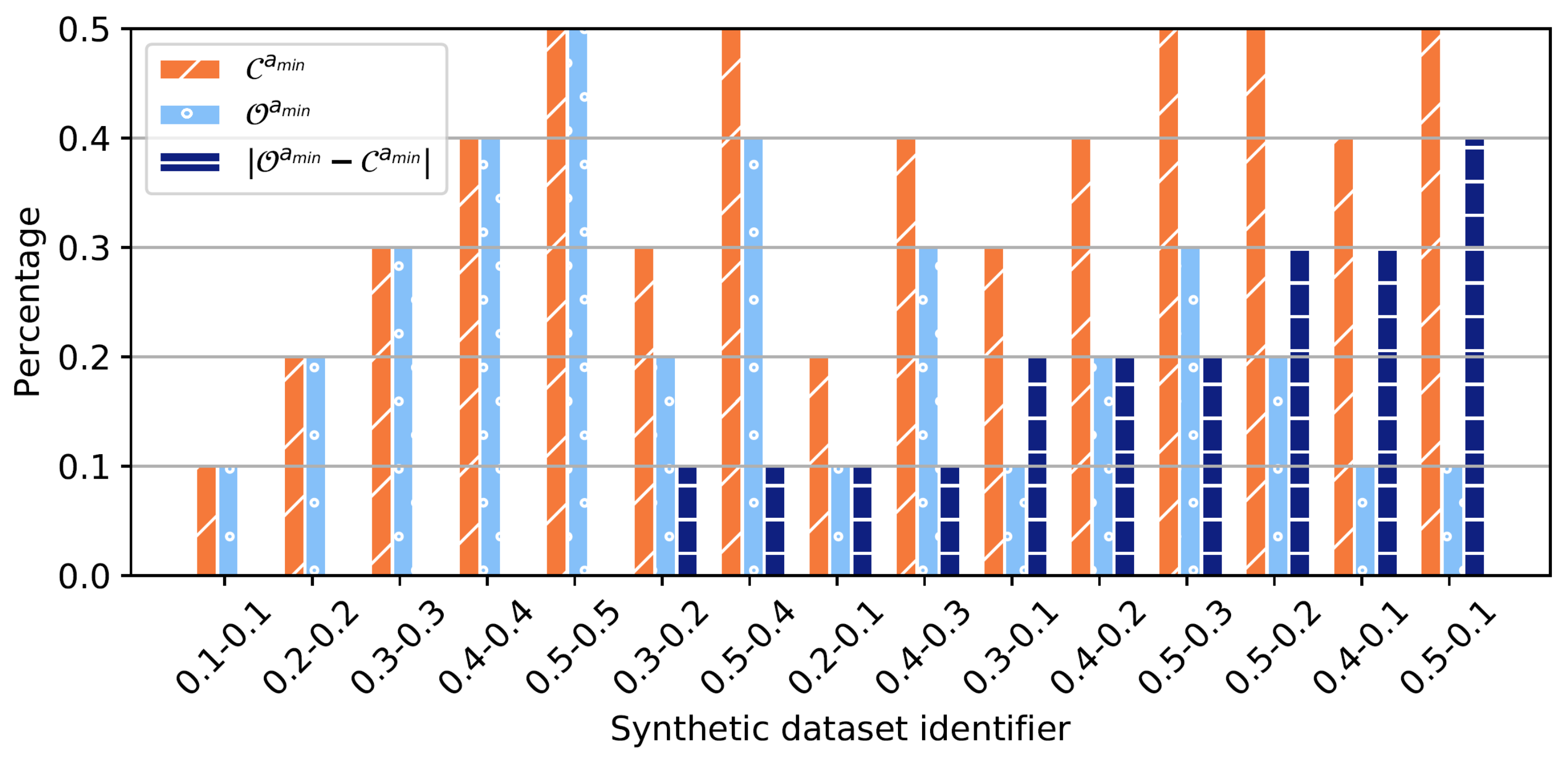}
    \caption{Provider Group Imbalance}
\end{subfigure}
\caption{\textbf{Synthetic Datasets Imbalance}. Popularity tail across items based on the observed interactions, conveyed by each of our synthetic dataset, according with the procedure in Appendix \ref{sec:syn-data} (a). Catalog and observation representations of the minority group in synthetic data, where C stands for ``Catalog", O stands for ``Interactions", and $\Delta$ C-O is the difference between catalog and interactions representations.}
\label{fig:item-rating-fairness}
\end{figure}

To assess the interplay among representations in the catalog, interactions, and relevance, we generate a range of synthetic datasets that simulate different catalog and observation imbalances. To create them, we use a procedure based on two stochastic block models \cite{DBLP:conf/nips/YaoH17}, whose description is provided in Appendix \ref{sec:syn-data}. The popularity tails, catalog, and observation representations of the resulting $15$ synthetic datasets are reported in  \figurename~\ref{fig:item-rating-fairness}. Through synthetic datasets, we explore a wider range of configurations, questioning situations not usually observable in public real-world datasets but that might occur in the real world, e.g., datasets with different representations of the minority group. 

\begin{figure}[!t]
\centering
\includegraphics[width=1.0\linewidth]{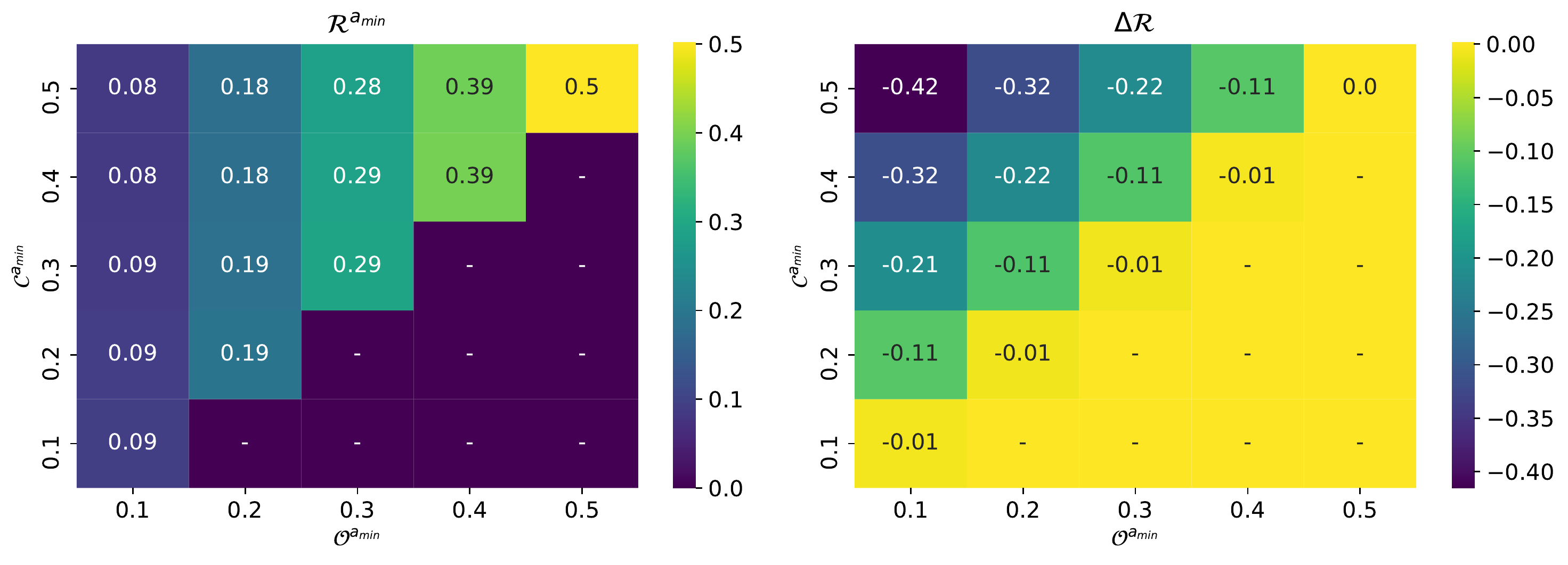}
\caption{\textbf{Contribution-Relevance Relationship}. The percentage of relevance given to the minority (left) and the difference between contribution and relevance percentages (right).}
\label{fig:synthetic-1}
\end{figure}

Once the synthetic datasets are generated, we run the pair-wise optimization procedure on all our synthetic datasets. Then, we analyze the resulting relevance scores for each provider group with respect to their contribution in the catalog and their representation in the interactions. To this end, \figurename~\ref{fig:synthetic-1} depicts the share of the items' relevance for the minority group (left) and the difference between contribution and relevance shares for the minority group (right). It should be noted that the half-lower diagonal of the heatmap is not considered, given that we only generate synthetic datasets where the difference in representation between contribution and interactions for the minority group is non negative. Results show us that the representation in relevance (left heatmap) is consistent across datasets having the same representation of the minority group in interactions, i.e., within the same column (e.g., 0.5-0.4 and 0.4-0.4 settings). Further, for each dataset, the relevance is similar to the representation in interactions and increases as much as the representation in interactions increases (from left to right). It follows that the representation in interactions for the minority group appears to play a key role in shaping the share of relevance for the group. By extension, the disparate relevance may directly depend on the gap between the representation in contribution and in interactions (right heatmap). The smaller the gap is, the lower the difference between (i) the representation of the minority group in the catalog and (ii) the share of relevance assigned to it is. The heatmaps allow us to see to what extent the imbalance between catalog and interaction representations influences the disparate relevance. We can draw the following observation.    

\hlbox{Observation 1}{The percentage of relevance for the minority depends on the difference between contribution and interaction representations. The larger the difference is, the larger the disparate relevance is.}

\begin{figure}[!t]
\begin{subfigure}[t]{0.48\linewidth}
    \centering
    \includegraphics[width=1.0\linewidth]{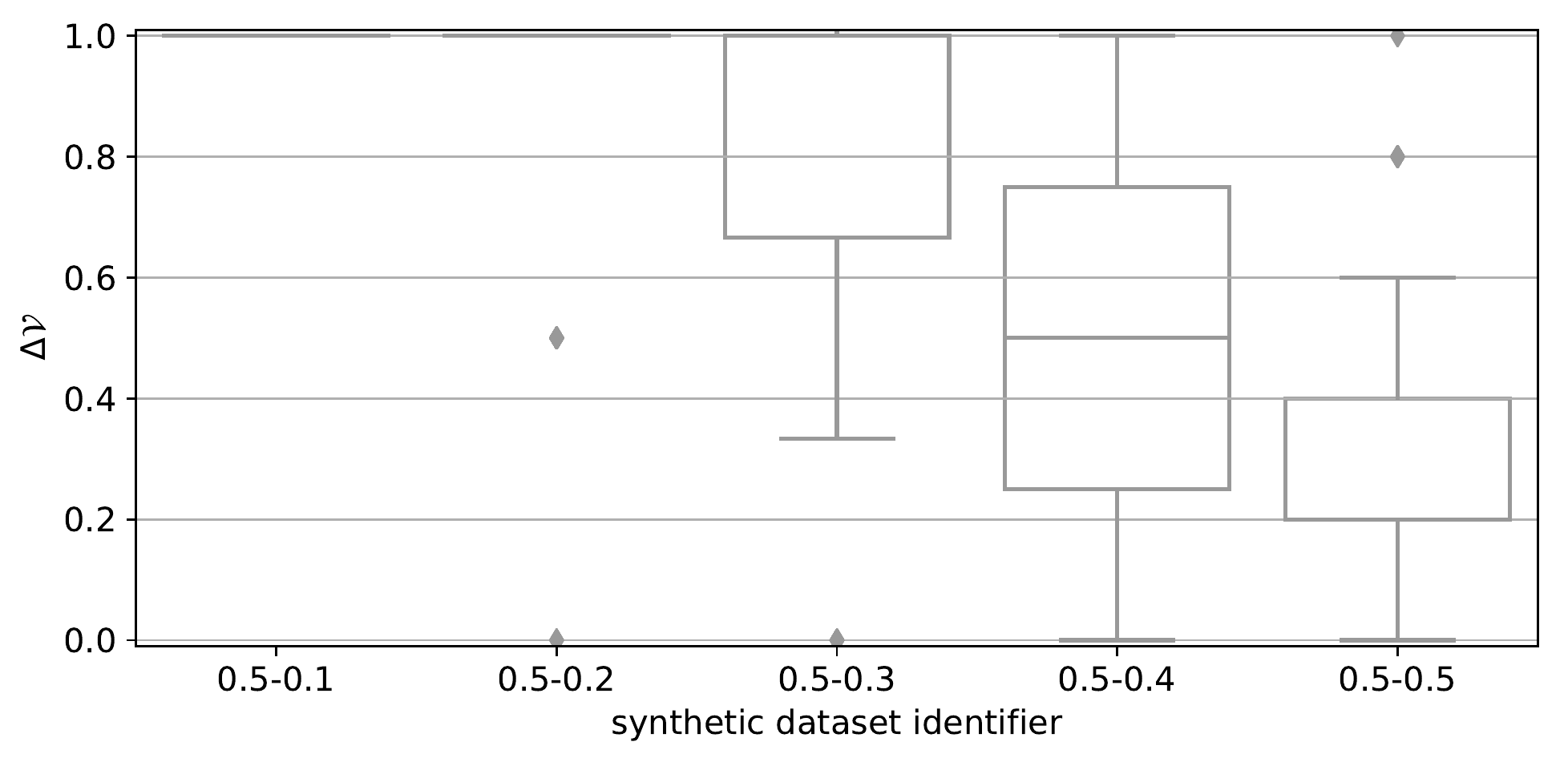}
    \caption{Disparate visibility for the minority}
    \label{fig:synthetic-2-1}
\end{subfigure}
\begin{subfigure}[t]{0.48\linewidth}
    \centering
    \includegraphics[width=1.0\linewidth]{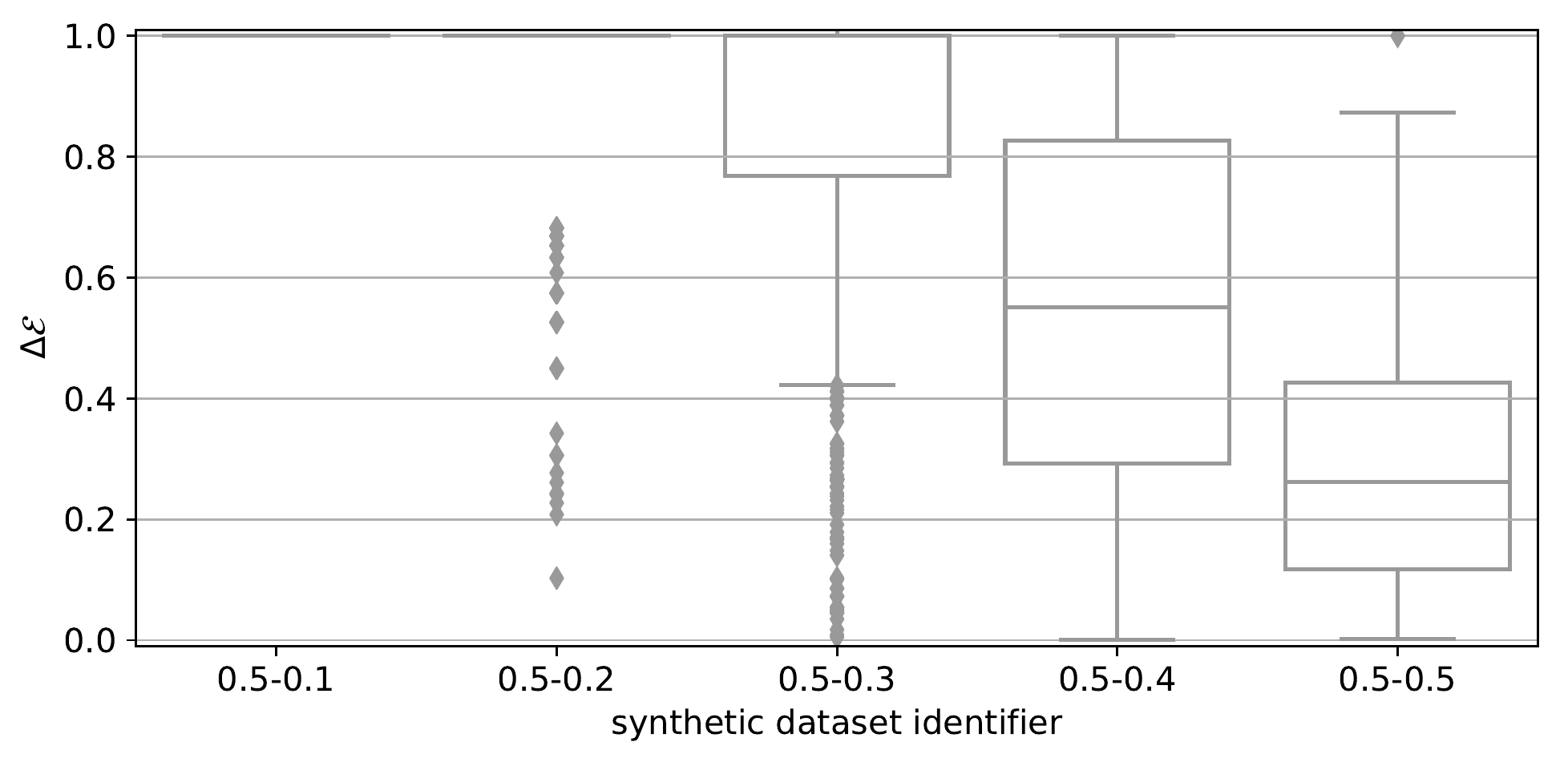}
    \caption{Disparate exposure for the minority}
    \label{fig:synthetic-2-2}
\end{subfigure}
\caption{\textbf{Disparate Impacts}. Disparate visibility (a) and exposure (b) for the minority group $\text{a}_{\min}$ in top-$10$ lists. The disparate impacts are calculated with Eq. \ref{eq:dv} and \ref{eq:de}, respectively. }
\label{fig:synthetic-2}
\end{figure}

Next, according to the relevance learnt by the recommendation model on each synthetic dataset, we suggested to each user $k=10$ items; then, in \figurename~\ref{fig:synthetic-2}, we measured the disparate visibility (exposure) for the minority group, both ranging between $[0, 1]$; we consider visibility as the percentage of providers of a given group in the recommendations (regardless of their position in the ranking), while we use a definition of exposure inspired by Singh and Joachims \cite{DBLP:conf/kdd/SinghJ18}. Both have been previously introduced in Section \ref{sec:disp-impacts}. The higher the value is, the higher the disparate impact is. The connection of all these results allows us to understand how much the imbalances in relevance for provider groups, learnt by recommender system, result in inequalities on recommended lists.    

\hlbox{Observation 2}{In contexts with high catalog-interaction imbalances, there is a larger disparate visibility (exposure) against the minority group, based on its contribution in the catalog. The higher the disparate relevance is, the higher the disparate visibility (exposure) is.}

We can observe that the effect on exposure is more evident. We conjecture that this result might depend on the fact that, when in presence of a small minority, the items from the minority group are progressively inserted at lower positions of the top-10 or even excluded, because of the lower predicted relevance. The considerations we made suggest to investigate treatments that impact on the interaction and relevance distributions. Hence, we will play with the minority group representation in interactions and regularize the percentage of relevance given to items across groups. 

\section{Reducing Disparities via Upsampling and Regularization} \label{sec:mitigation}
With an understanding of our fairness goals and of the intuitions we came up with in the exploratory study, this section describes how we can arrange a recommender system to reduce disparities, while preserving utility. 

Our exploratory analysis revealed that the share of relevance may depend on the representation of providers' groups in both the catalog and the interactions, and that the more similar the two representations are for a group, the lower the resulting disparate relevance is. It is unlikely that this property is met in interactions collected from real-world platforms, as we will later show. It follows that controlling the balance among catalog-interaction representations for a group could require to act on the interactions. To this end, we will up-sample interactions of the minority group, to reduce existing imbalances. 

Balanced representations of the minority group between the catalog and the interactions would not necessarily ensure a lower disparate relevance in real-world situations. Differently from the synthetic data we generated, interactions in the real world show several imbalances (e.g., due to presentation, preferences, user interfaces), which are hard to simulate, that may still distort the output relevance. It follows that, when an upsampling mechanism is not sufficient to accomplish our goals, we need a regularization approach to account for the distribution of relevance across groups during learning. Only regularizing the distribution of relevance across groups, with no upsampling, may not be enough as well, if minority interactions are too few. Hence, our treatment will control the interplay between upsampling and regularization.

To deal with upsampling, we play with the data sampling strategies that generate interaction instances (i.e., observed user-item pairs); conversely, to account for relevance, we will define a training loss function aimed to minimize the pair-wise error specified in Eq. \ref{eq:pair-wise} and the disparate relevance defined in Eq. \ref{eq:disp-rel}. We will show empirically that, although the optimization relies on a given set of interactions, even artificially up-sampled, the approach generalizes to real and unseen interactions. The treatment builds upon the following steps: 

\vspace{2mm} \noindent \textbf{Interaction Upsampling}. We propose to up-sample interactions related to the minority group with different user-item selection techniques, with the aim of covering a range of alternative setups:

\begin{itemize}[leftmargin=*]
\item \texttt{real} consists of an upsampling of existing interactions belonging to the minority group, with repetitions. Specifically, we select the item of the existing user-item interaction to be up-sampled, based on a probability function that takes into account the contribution of the minority $s_i^{\text{a}_{\min}}$, for each item $i$. The higher the contribution of the minority group, the higher the probability to be selected. Then, the real interactions involving the selected item $i$ are retrieved, and the one to be up-sampled is randomly selected.   
\item \texttt{fake} stands for a random upsampling on synthetic interactions, with no repetitions. This strategy adds new interactions related to items from the minority group. Similarly to \texttt{real}, the item involved in the up-sampled interaction is selected based on a probability function that accounts for the contribution of the minority $s_i^{\text{a}_{\min}}$, for each item $i$. Then, the user to be included in the up-sampled interaction is randomly selected among those users of $U$ who have not already interacted with item $i$.
\item \texttt{fake-by-pop} refers to an upsampling of synthetic interactions based on item popularity, with no repetitions. Given items with at least one provider from the minority, the item to be inserted in the up-sampled observation is selected according to an item-popularity probability. The higher the popularity is, the higher the probability to be selected is. The user of the up-sampled interaction is randomly chosen among those users of $U$ who have not already interacted with item $i$. 
\end{itemize}

These strategies assume to up-sample pairs $(u,i)$, until the representation of the minority group in the interactions meets a target percentage of the total interactions. This percentage, investigated in the experimental section, will first target the representation of the minority group in the catalog.  

\vspace{2mm} \noindent \textbf{Regularized Optimization}. Given a range of batches of training data samples $T_{batch}$ (i.e., either pairs for a point-wise approach or triplets for a pair-wise approach), built on top of the up-sampled interactions, each training batch is fed into a model that follows a regularized paradigm, derived from a traditional optimization setup. The loss function can be formalized as follows:     

\begin{equation}
\underset{\theta}{\operatorname{argmax}} \,\, (1 - \lambda) \, \text{acc}(\text{T}_{\text{batch}}) - \lambda \, \text{reg}(\text{T}_{\text{batch}}) 
\label{eq:correlation}
\end{equation}

\vspace{2mm}

\noindent where $acc(T_{batch})$ is the original accuracy loss, computed over $T_{batch}$. In our experimental study, we deal with a pair-wise optimization, thus the accuracy loss is computed as in Eq. \ref{eq:pair-wise}. The $\lambda\in[0,1]$ parameter expresses the trade-off between accuracy and disparate relevance. With $\lambda=0$, we yield the output of the recommender, not taking disparate relevance into account. Conversely, with $\lambda=1$, the output of the recommender is discarded, and we focus on minimizing disparate relevance.

The regularization term, $reg(T_{batch})$, operationalizes our strategy of disparate relevance minimization, based on Eq. \ref{eq:disp-rel}. The proposed criterion is equivalent to compute, in percentage, the relevance received by minority-group items in a batch with respect to the total relevance received by all items in that batch, and then balance it to the percentage of contribution of the minority group in the catalog. Let $C^{\text{a}_{\min}}$ be the contribution of the minority group in the catalog, computed as in Eq. \ref{eq:contribution}, the regularization can be defined as follows: 

\begin{equation}
\begin{split}
reg(\text{T}_{\text{batch}}) = \left( \frac{\sum_{(u,i,\_) \in \text{T}_{\text{batch}}} f_{\theta}(u,i) \cdot S_i^A({\text{a}_{\min}})}{\sum_{(u,i,\_) \in \text{T}_{\text{batch}}} f_{\theta}(u,i)} - C^{\text{a}_{\min}} \right)^2
\end{split}
\label{eq:regularization}
\end{equation}

\vspace{2mm}

\noindent where $S_i^A({\text{a}_{\min}}) = s_i^A({\text{a}_{\min}}) / \sum_{a \in A} s_i^A(a)$ is the percentage of minority providers who have been involved in the production/creation of item $i$. These regularized optimization implies that the model is penalized if the difference in relevance and contribution for the minority group of providers is high. The choice of the squared value, instead of an L2 norm or an Earth mover distance as examples, has been proved to be of benefit in optimization, while being simple and effective. Our framework can be easily extended to other options. The contextualization with respect to the literature is presented in Section~\ref{sec:conn-treat}.

\section{Experimental Treatment Evaluation and Analysis} \label{sec:experimental-evaluation}
In this section, we empirically study the effects of each component of our treatment and of the treatment as a whole on the needs of both users (i.e., recommendation utility) and providers (i.e., disparate relevance, visibility, and exposure). We answer the following four research questions: 

\vspace{1mm} \noindent \textbf{RQ1.} How much should we up-sample minority-group interactions to improve the trade-off between recommendation utility and disparities?  

\vspace{1mm} \noindent \textbf{RQ2.} To what extent do upsampling and regularization impact on the trade-off between recommendation utility and disparities, individually and jointly?

\vspace{1mm} \noindent \textbf{RQ3.} How does our treatment concretely reduce disparities for the minority group? How does it impact on internal mechanisms?

\vspace{1mm} \noindent \textbf{RQ4.} To what extent does our treatment affect disparities, utility, and coverage, compared with others? Can the latter benefit from regularized relevances?

\subsection{Experimental Setup} 

\subsubsection{Datasets}\label{sec:dataset}
In order to validate and ensure the reproducibility of our proposal, we selected datasets that are publicly available, covering different domains. We remark that this experimentation is made difficult because there are very few datasets targeting our scenario, and the  datasets we consider are highly sparsed.

\vspace{1mm} \noindent \textbf{Movielens-10M (ml-10m)}~\cite{DBLP:journals/tiis/HarperK16} includes $10M$ ratings applied to $10k$ movies by $72k$ users. In order to be fed into a pair-wise model, interactions are binarized using a threshold (i.e., ratings equal or higher than $3$ are marked as $1$, the other ones are changed to $0$). This dataset does not contain sensitive attributes of the providers and there is no notion of provider. Our study considers movie directors as providers to reflect a real-world scenario. To link movies to their corresponding directors, we capitalized on the methods offered by the TMDB APIs\footnote{\url{https://developers.themoviedb.org/3}}. Specifically, we used the \emph{getCredits(tmdbId)} method to retrieve data about people involved in the movie\footnote{Please note that the links.csv file in Movielens includes movieId-tmdbId associations.}. We filtered records for individuals with ``\emph{Director}" as a role. Then, we called the \emph{getDetails(peopleId)} method, passing the id retrieved for each director. The latter method outputs a list with the name and the gender of the director. Note that there are movies with more than one director. The representation of women directors is around $6\%$ in the catalog and $3.9\%$ in the interactions.

\vspace{1mm} \noindent \textbf{COCO Course Collection (coco)}~\cite{DBLP:conf/worldcist/DessiFMR18} includes $74k$ learners, who gave $600k$ ratings to $10k$ online courses. Similarly to \emph{ml-10m}, ratings are binarized using a threshold (i.e., ratings equal to $5$ are marked as $1$, the other ones are changed to $0$). We selected this threshold due to the extremely high imbalance among rating values, as reported in the original paper. In this scenario, we assume that instructors act as providers. Providers representing a company or an institution were removed, since there was no practical way to associate their items to gender representations. One or more instructors could cooperate in the same course. However, no information about their gender is reported. To extract this attribute, we considered their naming information\footnote{We point out the challenges seeking to include genders determined by naming information, considering that the retrieved gender might not match the expected gender for someone. Related to that issue is the problem of the assumption of a binary gender. Most datasets and tools only consider two genders, ``male'' and ``female'', so we have no chance to also consider non-binary attributes. While keeping this in mind, we recognize all genders should be respectfully treated and our framework naturally adapts to multi-class attributes and non-binary genders; we believe that our study will deserve attention in this context.}. Specifically, we used the methods offered by GenderAPIs\footnote{\url{https://gender-api.com/}}, that allow to determine the gender by naming information, with a certain confidence. Such a practice has been conducted in prior work to deal with the absence of gender labels~\cite{DBLP:conf/chi/ChenMHW18,DBLP:conf/recsys/MansouryMBP19}. Only predictions with a confidence higher than $75\%$ were kept. The representation of women instructors in the catalog is around $17\%$, reduced to $12\%$ in the interactions.

\subsubsection{Evaluation Metrics} \label{sec:metrics}
In this section, we present the metrics we considered to assess the impact of our work. In addition to the disparity metrics introduced in Section \ref{sec:disp-impacts}, which cover the aspects associated with providers' fairness, several other perspectives of the recommender system should be considered. Our study in this paper also includes an assessment (i) of personalization in terms of recommendation utility, and (ii) of coverage of items for the provider groups and as a whole.   

\vspace{2mm} \noindent \textbf{Personalization}. To evaluate personalization, we compute the utility of recommended lists via \emph{Normalized Discounted Cumulative Gain} (NDCG)~\cite{DBLP:journals/tois/JarvelinK02}.

\begin{equation}
\text{DCG}(k|\theta) = \sum_{u \in U} \widetilde{R}_{u,\rho_{\theta}(u,1)} + \sum_{pos=2}^{k} \frac{\widetilde{R}_{u,\rho_{\theta}(u,pos)}}{log_2(pos)}
\end{equation}

\begin{equation}
\text{NDCG}@k(k|\theta) = \frac{\text{DCG}(k|\theta)}{\text{IDCG}(k|\theta)}
\label{eq:ndcg}
\end{equation}

\vspace{2mm}

\noindent where $\rho_{\theta}(u,pos)$ is the item $i$ recommended to user $u$ at position $pos$, and the values in $\widetilde{R}$ formalized in Section~\ref{sec:recsys-formalization} are considered as user-item relevances, while computing $DCG$. The ideal $DCG$ is calculated by sorting items based on decreasing true relevance (i.e., for an item, the true relevance is $1$ if the user interacted with the item in the test set, $0$ otherwise). The higher the NDCG score achieved by the recommender system is, the more effective the generated recommendations are for consumers. 

\vspace{2mm} \noindent \textbf{Item Coverage}. In addition to personalization and disparate impacts, we measure the total coverage of items ($\text{Cov}_{\text{tot}}$) and of items delivered by providers in the minority ($\text{Cov}_{\text{a}_{\min}}$) and the majority ($ \text{Cov}_{\overline{\text{a}_{\min}}}$) group. Coverage is an important property \cite{DBLP:journals/tiis/KaminskasB17}, since an approach that only increases the recommendation of one item provider of the minority group would not likely fair within the minority group. 

\begin{equation}
    \text{Cov}_{\text{tot}} = \frac{1}{|I|} \sum_{i \in I} \min\left(1, \sum_{u \in U} |\widetilde{I}_u \cap \{i\}|\right)
\end{equation}
\vspace{1mm} 
\begin{equation}
    \text{Cov}_{\text{a}_{\min}} = \frac{1}{|I^{\text{a}_{\min}}|} \sum_{i \in I^{\text{a}_{\min}}} \min\left(1, \sum_{u \in U} |\widetilde{I}_u \cap \{i\}|\right)
\end{equation}
\vspace{1mm} 
\begin{equation}
    \text{Cov}_{\overline{\text{a}_{\min}}} = \frac{1}{|I \setminus I^{\text{a}_{\min}}|} \sum_{i \in I \setminus I^{\text{a}_{\min}}} \min\left(1, \sum_{u \in U} |\widetilde{I}_u \cap \{i\}|\right)
\end{equation}

\vspace{2mm} \noindent where $I^{\text{a}_{\min}} = \{i : s_i^A(\text{a}_{\min})>0\}$ is the set of items that have at least one provider belonging to the minority group. Each coverage score ranges in $[0,1]$, with values close to $1$ for higher coverage values.

\subsubsection{Experimental Setting}
We considered several optimization settings, each one characterized by a different combination of upsampling and regularization treatments, as proposed in Section \ref{sec:mitigation}. They are briefly identified as follows:
\begin{itemize}[leftmargin=*]
\item \texttt{baseline}: training without any upsampling and regularization treatment;
\item \texttt{real}: only real upsampling;
\item \texttt{fake}: only fake upsampling;
\item \texttt{fake-by-pop}: only fake-by-pop upsampling;
\item \texttt{reg}: only regularization;
\item \texttt{real+reg}: real upsampling, followed by regularization;
\item \texttt{fake+reg}: fake upsampling, followed by regularization;
\item \texttt{fake-by-pop+reg}: fake-by-pop upsampling, followed by regularization.
\end{itemize}

\subsubsection{Implementation Details}
For each dataset, a temporal train-test split was performed by including the last $20\%$ of interactions released by a user into the test set, $10\%$ of interactions were included into the validation set, and the remaining $70\%$ oldest ones into the training set~\cite{DBLP:journals/umuai/CamposDC14,SanchezB20}. Embedding matrices, with vectors of size $100$, were initialized with values uniformly distributed in $[0, 1]$. The optimization function was transformed to the equivalent minimization dual problem. During training, the model was served with batches of $1,024$ training triplets, chosen from a pre-computed set of triplets. To populate it, for each user $u$, we create $10$ triplets $(u, i, j)$ per observed item $i$; the unobserved item $j$ is randomly selected for each triplet. Before each epoch, we shuffle the training batches. The learning rate for the \emph{Adam} optimizer is $0.01$. The $dot$ function was used to compute the similarity (i.e., the relevance) between user and item vector. Each model was trained until convergence on the validation set, for a maximum of $100$ epochs. 

\subsection{Experimental Results}

\begin{figure}[!b]
\begin{subfigure}[t]{0.32\linewidth}
    \centering
    \includegraphics[width=1.0\linewidth]{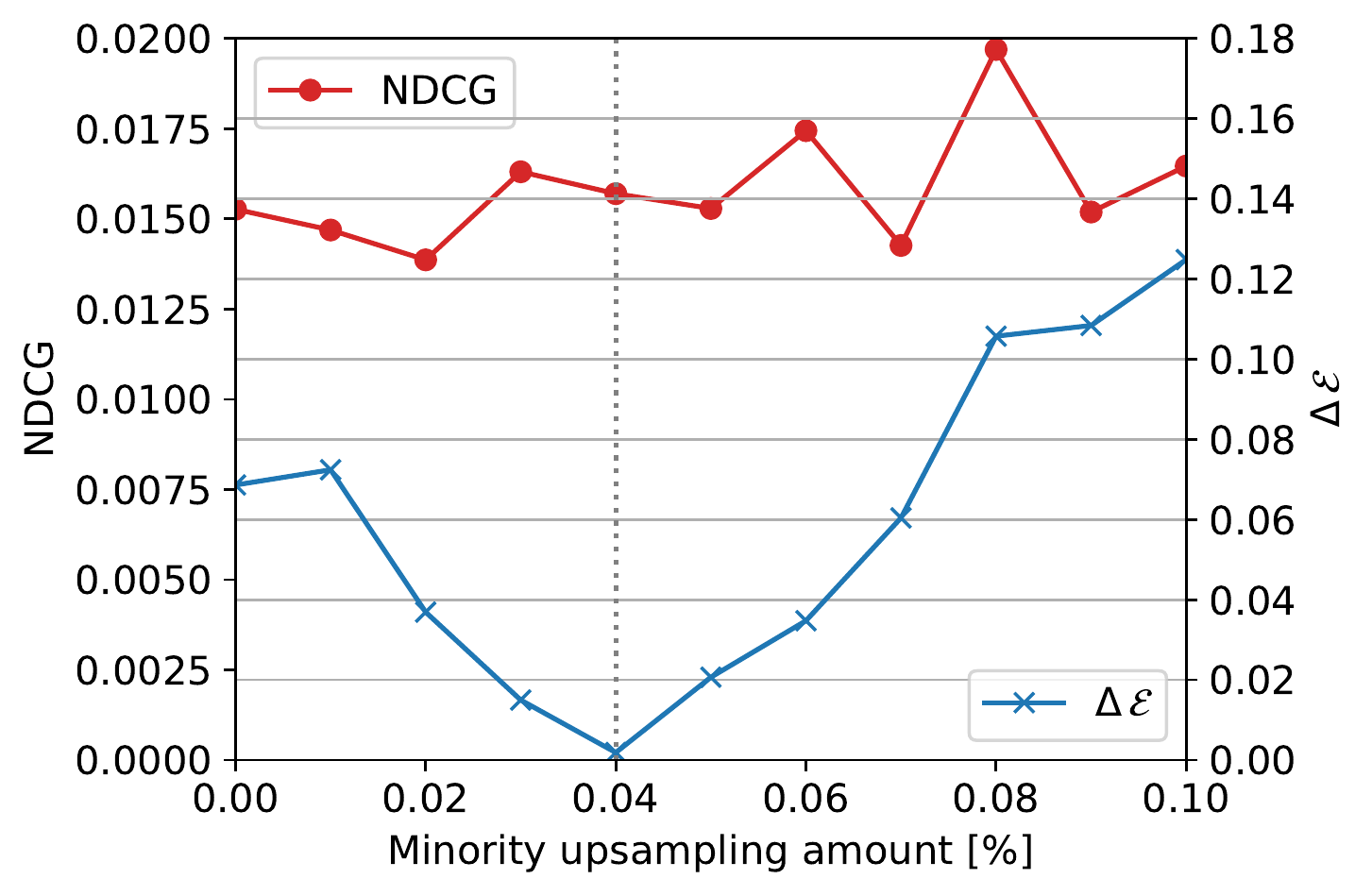}
    \caption{Real on coco}
    \label{fig:coco-1-1}
\end{subfigure}
\begin{subfigure}[t]{0.32\linewidth}
    \centering
    \includegraphics[width=1.0\linewidth]{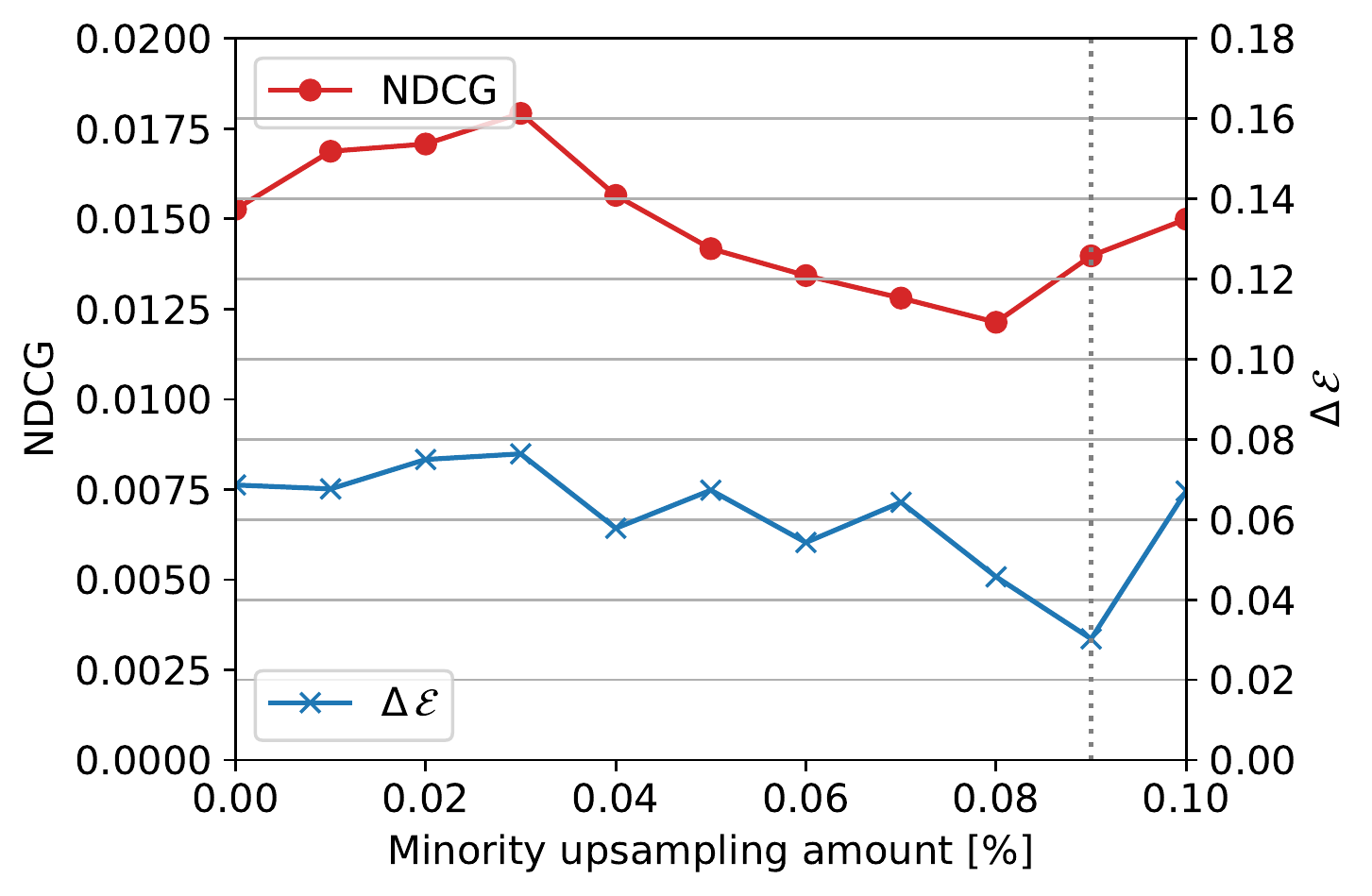}
    \caption{Fake on coco}
    \label{fig:coco-1-2}
\end{subfigure}
\begin{subfigure}[t]{0.32\linewidth}
    \centering
    \includegraphics[width=1.0\linewidth]{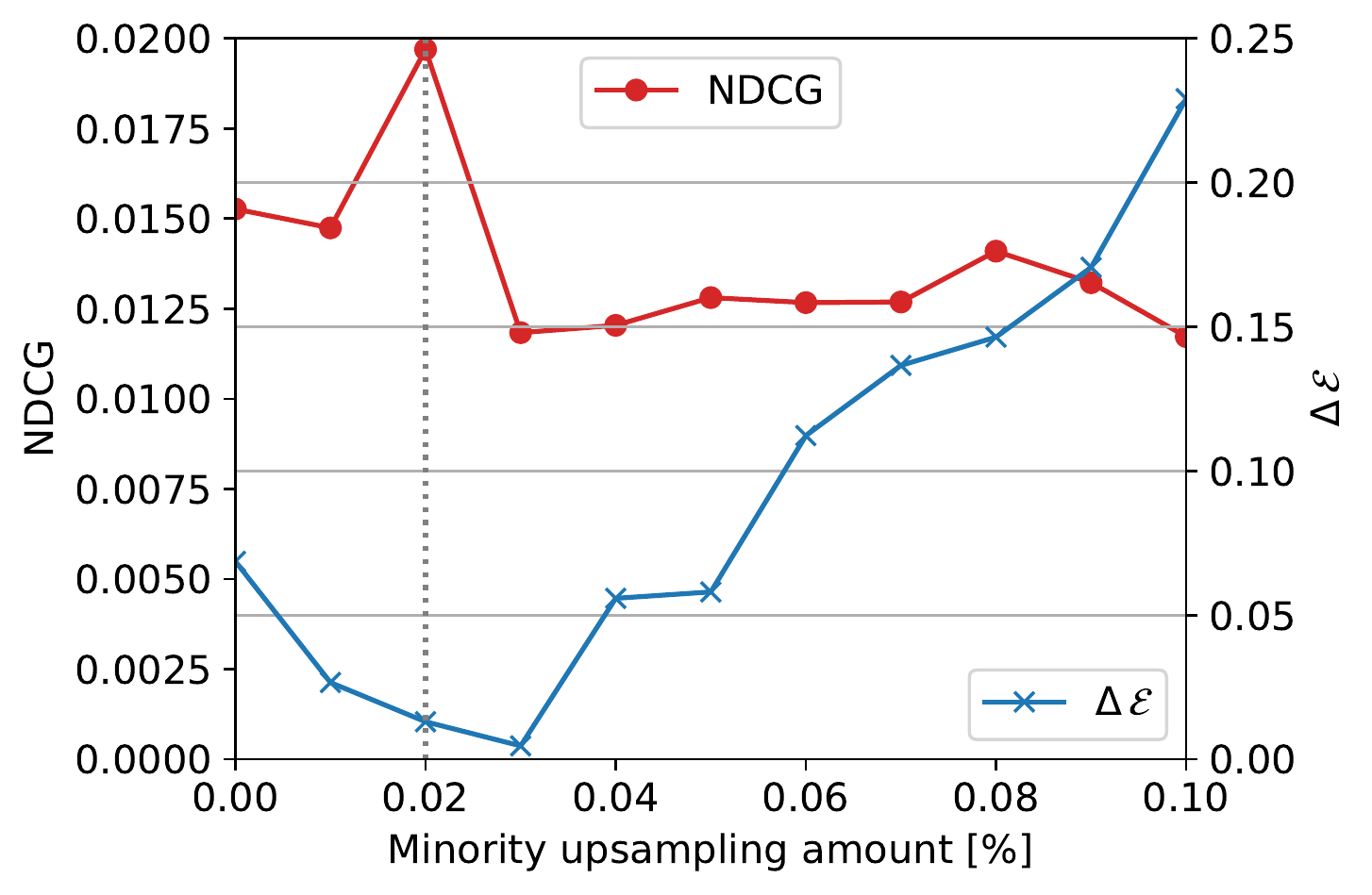}
    \caption{Fake-by-pop on coco}
    \label{fig:coco-1-3}
\end{subfigure}
\begin{subfigure}[t]{0.33\linewidth}
    \centering
    \includegraphics[width=1.0\linewidth]{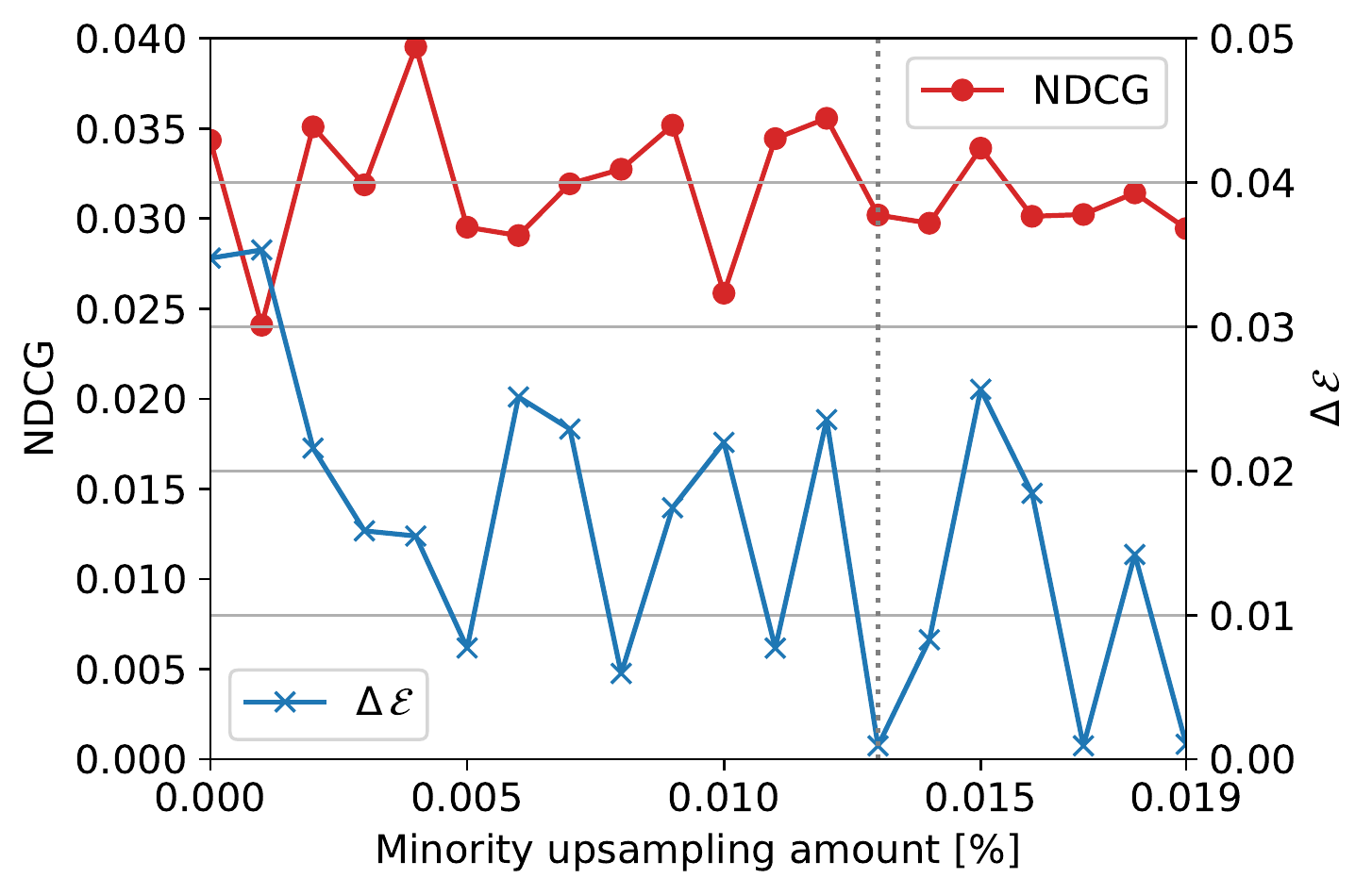}
    \caption{Real on ml-10m}
    \label{fig:ml-10m-1-1}
\end{subfigure}
\begin{subfigure}[t]{0.32\linewidth}
    \centering
    \includegraphics[width=1.0\linewidth]{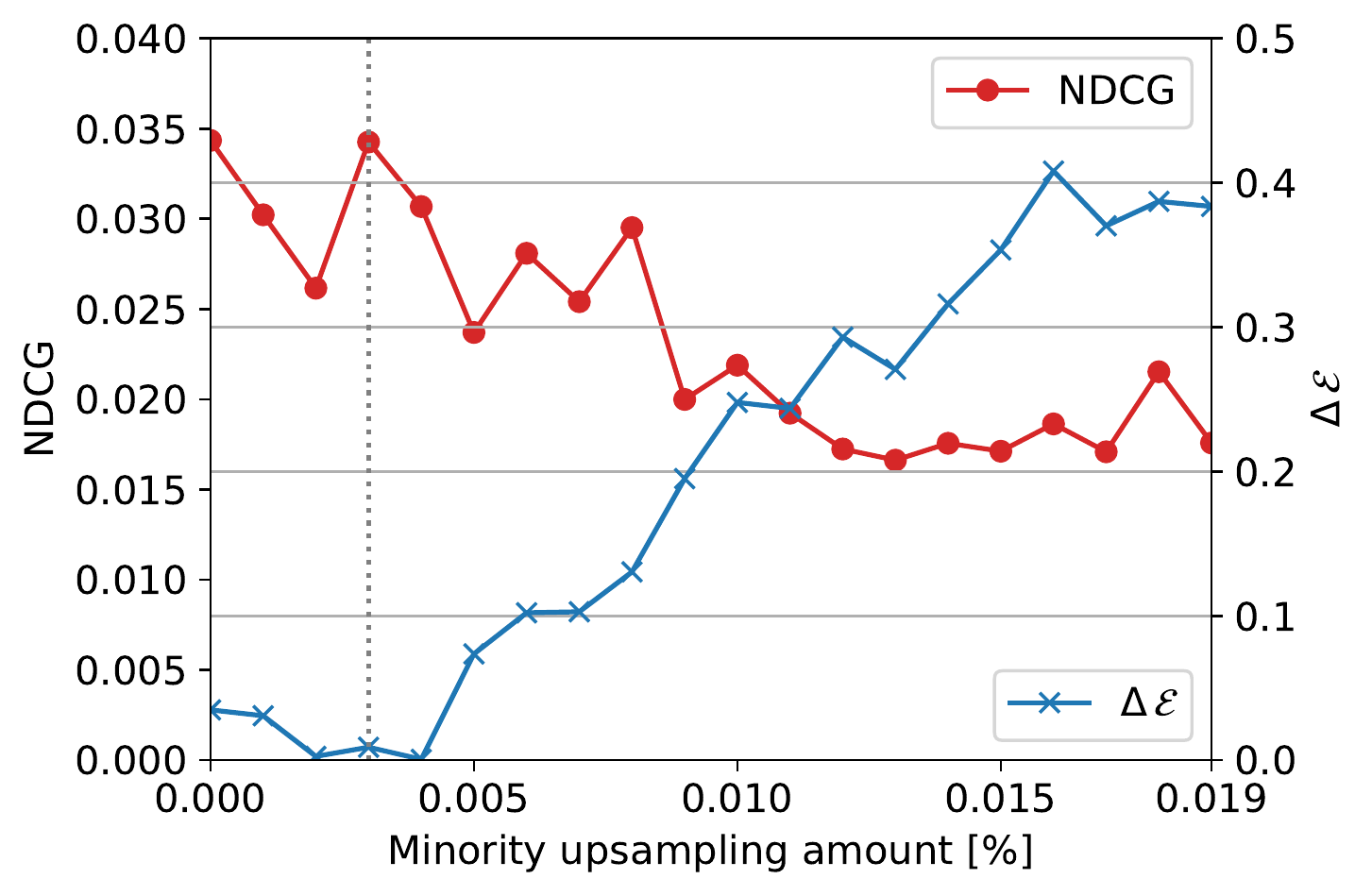}
    \caption{Fake on ml-10m}
    \label{fig:ml-10m-1-2}
\end{subfigure}
\begin{subfigure}[t]{0.33\linewidth}
    \centering
    \includegraphics[width=1.0\linewidth]{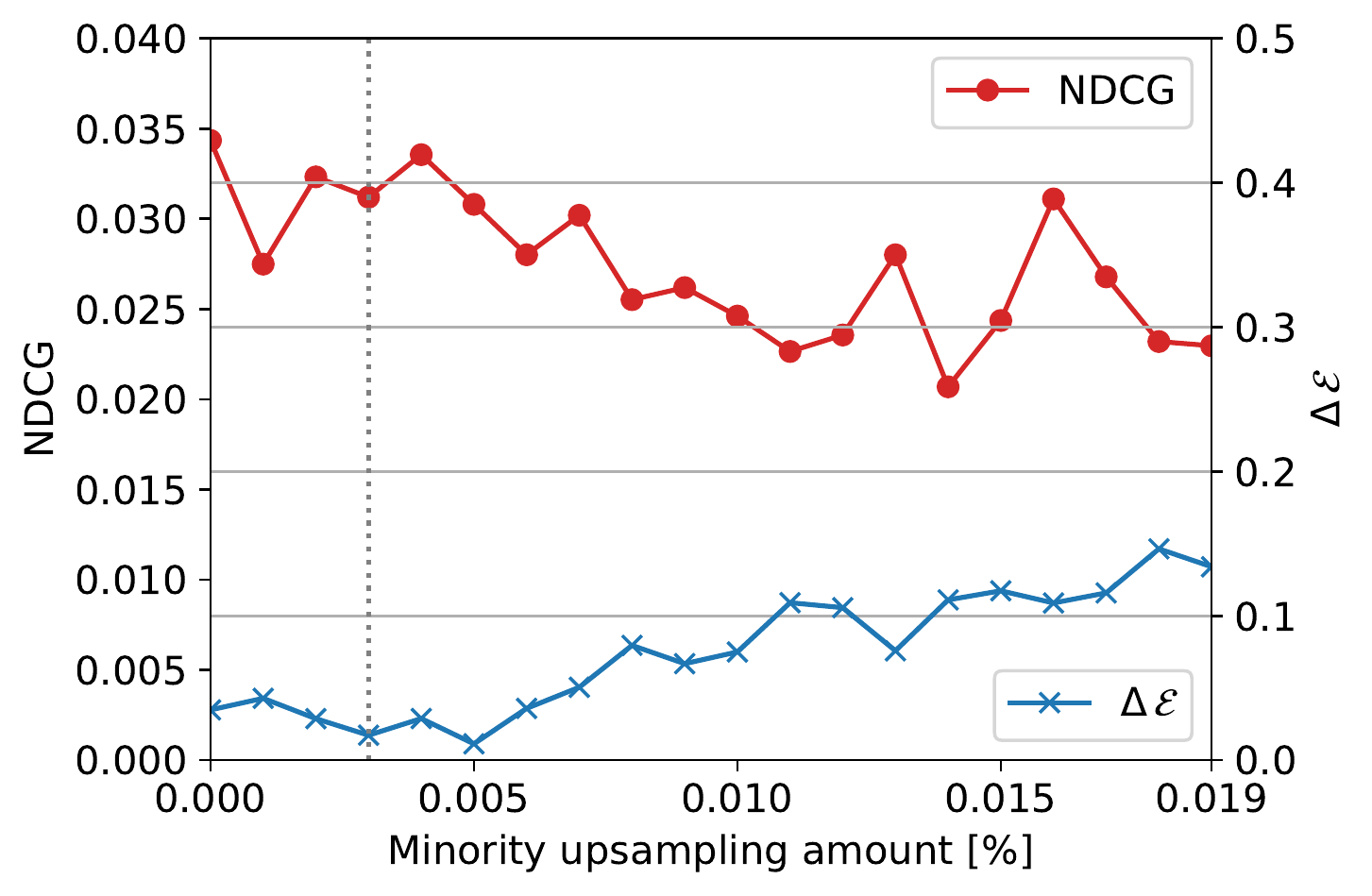}
    \caption{Fake-by-pop on ml-10m}
    \label{fig:ml-10m-1-3}
\end{subfigure}
\caption{\textbf{Influence of Upsampling Degree on Trade-off}. The trade-off between Normalized Discounted Cumulative Gain (NDCG: red line with bullet markers) and Disparate Exposure ($\Delta \, \mathcal{E}$: blue line with star markers) based on the degree of upsampling, varying the upsampling techniques and datasets. Dotted lines indicate the degree of upsampling resulting in a good trade-off (i.e., high NDCG and low $\Delta \, \mathcal{E}$). Disparate visibility and relevance showed similar patterns and are omitted for the sake of clarity and readability.}
\label{fig:rq1}
\end{figure}

\subsubsection{Comparing Upsampling Techniques (RQ1)} 
With this experiment, we aim to understand to what degree upsampling influences recommendation utility and disparate impacts on group relevance, on visibility, and on exposure, and investigate how and how much we should up-sample to obtain a good trade-off among the metrics. Although our exploratory study revealed that paring the percentage of interactions for the minority group with the percentage of contribution in the catalog may be the best choice, interactions in real world show several imbalances that may distort the output relevance. Hence, we experiment with different degrees of upsampling, not just targeting a minority-group representation in the interactions equal to its representation in the catalog. 

To this end, for each dataset and upsampling technique, we created a range of model instances fed with a different amount of up-sampled data, using the upsampling techniques described in Section \ref{sec:mitigation}. Results in Figure \ref{fig:rq1} depict NDCG and $\Delta \mathcal{E}$ at increasing percentage of minority observation upsampling. Patterns related to $\Delta \mathcal{R}$ and $\Delta \mathcal{V}$ were similar to the ones obtained on $\Delta \mathcal{E}$, so we do not report them for conciseness and readability. The considered plots show us that NDCG tended to decrease, when the amount of up-sampled data became larger. The loss in recommendation utility depends on the dataset and the technique, with \texttt{fake} suffering from the largest loss. Conversely, we observed that $\Delta \, \mathcal{E}$ achieved the lowest value for an upsampling between $15\%$-$20\%$, depending on the dataset. This latter behavior came from the fact that, for small upsampling amounts, the model tended to show a disparate impact in favor of the majority group. Increasing upsampling leads the minority to get more and more exposure; this can get to the point where the majority is affected by a disparate impact, i.e., the minority group is favored more than expected (e.g., in Figure \ref{fig:coco-1-1}, when upsampling is greater than $4\%$). 

Moving to the comparison of the results with different datasets, coco experiences a lower loss in NDCG for the same upsampling technique against ml-10m. Interestingly, for small upsampling amounts, NDCG ends up increasing in coco, with respect to the baseline, which does not make use of upsampling. Furthermore, coco is more susceptible to the amount of upsampling, resulting in larger variations of $\Delta \mathcal{E}$. Considering the same dataset and observing patterns for different upsampling techniques, it can be observed that \texttt{real} preserves a good level of NDCG, even for high amounts of upsampling.  Conversely, $\Delta \, \mathcal{E}$ follows similar patterns for all the upsampling techniques. An exception is made for \texttt{real} on ml-10m, which showed a decreasing while noisy trend on $\Delta \mathcal{E}$. Therefore, while upsampling in general is beneficial for controlling $\Delta \, \mathcal{E}$, each of the techniques differently preserves the NDCG originally achieved, changing the trade-off between effectiveness and disparate impacts.     

\hlbox{Observation 3}{The upsampling of minority-group interactions reduces disparate impacts, i.e., the inequality of exposure, visibility, and relevance with respect to the contribution of the minority group in the catalog. The loss in recommendation utility is negligible or even absent in many cases. The amount of needed upsampling depends on the dataset and the upsampling technique.}

\begin{table}[!t]
\resizebox{\textwidth}{!}{%
\begin{tabular}{cc|r|rrr|rrr}
\hline 
\textbf{Data} & \textbf{Type} & \textbf{NDCG} & \textbf{$\Delta \, \mathcal{R} $} & \textbf{$\Delta \, \mathcal{V}$} & \textbf{$\Delta \, \mathcal{E}$} & \textbf{$\text{Cov}_{\text{tot}}$} & \textbf{$\text{Cov}_{\text{a}_{\min}}$} & \textbf{$\text{Cov}_{\overline{\text{a}_{\min}}}$} \\
 \hline
\multirow{4}{*}{coco} & baseline & 0.0153 \,\; & 0.0770 \,\; & 0.0733 \,\; & 0.0686 \,\; & 0.2165 & 0.1413 & 0.2321 \\
 & real & 0.0157 \,\; & \textbf{0.0067} * &  \textbf{0.0077} * & \textbf{0.0018} * & \textbf{0.2523} & \textbf{0.2906} & 0.2443 \\
 & fake & 0.0140 * &  0.0347 * & 0.0351 * & 0.0302 * & 0.2494 & 0.2504 & \textbf{0.2491} \\
 & fake-by-pop & \textbf{0.0197} * & 0.0231 * & 0.0243 * & 0.0129 * & 0.2202 & 0.1444 & 0.2361 \\
 \hline
 \multirow{4}{*}{ml-10m} & baseline & \textbf{0.0344} \,\; & 0.0253 \,\; & 0.0361 \,\; & 0.0347 \,\; & 0.1654 & 0.1224 & 0.1682 \\
 & real & 0.0302 * & \textbf{0.0037} * & \textbf{0.0047} * & \textbf{0.0009} * & \textbf{0.1734} & 0.1776 & \textbf{0.1732} \\
 & fake & \textbf{0.0343} \,\; & 0.0085 * & 0.0077 * & 0.0088 * & 0.1725 & \textbf{0.1879} & 0.1715 \\
 & fake-by-pop & 0.0336 * & 0.0188 * & 0.0163 * & 0.0171 * & 0.1638 & 0.1069 & 0.1675 \\
 \hline
\end{tabular}}
\caption{\textbf{Impact of Upsampling on Recommended Lists}. Normalized Discounted Cumulative Gain (NDCG); Disparate Relevance ($\Delta \, \mathcal{R}$), Disparate Visibility ($\Delta \, \mathcal{V}$) and Disparate Exposure ($\Delta \, \mathcal{E}$) based on minority contribution in the catalog; Coverage of the catalog ($\text{Cov}_{\text{tot}}$), of items from $\text{a}_{\min}$ ($\text{Cov}_{\text{a}_{\min}}$) and of items from  $\overline{\text{a}_{\min}}$ ($\text{Cov}_{\overline{\text{a}_{\min}}}$). For each setting, we report results for the upsampling levels identified with dotted lines in \figurename~\ref{fig:rq1}.  Bold values refers to the best value across algorithms for a given dataset. (`*') indicates scores statistically different with respect to the baseline ($p=0.05$).}
\label{tab:upsampling-1}
\end{table}

To characterize the peculiarities of each upsampling technique, Table \ref{tab:upsampling-1} reports information on recommendation utility, disparate impact, and coverage for representative settings, which achieved a good trade-off. Results show us that, in general, upsampling brings benefits to disparate impacts and coverage, while preserving recommendation utility. Specifically, on coco, \texttt{real} experienced a disparate impact lower than $1\%$ at all levels (i.e., relevance, visibility, exposure) and doubles the coverage of minority-group items (i.e., column $\text{Cov}_{\text{a}_{\min}}$). Conversely, \texttt{fake-by-pop} allowed us to improve the original recommendation utility, but disparate impact and coverage did not experience the same gains of \texttt{real}. On ml-10m, similar patters were observed for \texttt{real}, even though the loss in NDCG was larger. Compared with coco, \texttt{fake} and \texttt{fake-by-pop} achieved a better trade-off among metrics on ml-10m.

\hlbox{Observation 4}{Upsampling real existing interactions involving the minority (\texttt{real}) can make it possible to achieve a good trade-off among recommendation utility, disparate impacts, and coverage. Upsampling minority-group interactions via fake user-item interactions (\texttt{fake} and \texttt{fake-by-pop}) is suitable when the minority group is very small.}

\subsubsection{Benchmarking Combined Treatments (RQ2)} 
Even though upsampling made it possible to achieve good trade-offs, there are still disparities that should be reduced. Hence, in this experiment, we are interested in understanding the impact of regularization on the representative settings considered in the previous section. To this end, we applied the regularization described in Section \ref{sec:mitigation} to each of the settings reported in Table \ref{tab:upsampling-1}. Given that the disparate impacts to get reduced are often small, we adopt a $\lambda=1e^{-6}$ as a regularization weight. Our empirical results with lower or larger $\lambda$ values led to unreasonable variations in the validation set. 

Results in Table \ref{tab:upsampling-2} show us recommendation utility, disparate impact, and coverage achieved by the model instance trained with upsampling and regularization jointly. When comparing results between \texttt{baseline} and \texttt{reg}, it can be observed that a plain regularization, without upsampling, fails to bring a proper reduction of disparate impact. This is caused by the fact that  the regularization depends on the amount of minority-group interactions, and the amount of such a data is small when upsampling is not performed. Conversely, the regularization can introduce benefits for the other settings, especially for \texttt{fake} and \texttt{fake-by-pop} settings. We can draw the following observation. 

\hlbox{Observation 5}{Combining regularization and upsampling is crucial to fine-tune trade-offs achieved with the upsampling-only instance, especially when the up-sampled user-item interactions are fake.}

The regularization is essential to fine-tune the trade-off in cases where the upsampling alone does not allow to reduce it anymore. On both coco and ml-10m, this effect is observed for the \texttt{fake} and \texttt{fake-by-pop}. With a small loss in NDCG, disparate impact and coverage experienced substantial improvements. Under the \texttt{real} scenario, the regularization improves NDCG, with a small loss in the other metrics. Each upsampling technique, combined with regularization, leads to a good trade-off between utility and disparity. 
 
\begin{table}[!t]
\resizebox{\textwidth}{!}{%
\begin{tabular}{cc|r|rrr|rrr}
\hline 
\textbf{Data} & \textbf{Type} & \textbf{NDCG} & \textbf{$\Delta \, \mathcal{R} $} & \textbf{$\Delta \, \mathcal{V}$} & \textbf{$\Delta \, \mathcal{E}$} & \textbf{$\text{Cov}_{\text{tot}}$} & \textbf{$\text{Cov}_{\text{a}_{\min}}$} & \textbf{$\text{Cov}_{\overline{\text{a}_{\min}}}$} \\
 \hline
\multirow{8}{*}{coco} & reg & 0.0182 * & 0.0668 * & 0.0666 * & 0.0659 * & 0.2581 & 0.1965 & 0.2708 \\ 
&  (gain/loss)  & +18.95\% \; & -13.24\% \; & -9.14\% \; & -3.93\% \; & +19.21\% & +39.06\% & +16.67\% \\
 \cline{2-9} 
 & real+reg & 0.0178 * & 0.0134 * &  0.0148 * & 0.0087 * & 0.2425 & \textbf{0.2728} & 0.2357 \\ 
 & (gain/loss) & +13.37\% \; & $\ge$ +100\%   & +92.20\% \; & $\ge$ +100\%  & -3.88\% & -6.12\% & -3.52\% \\
 \cline{2-9} 
 & fake+reg & 0.0136 \,\; &  \textbf{0.0044} * & \textbf{0.0065} * & 0.0102 * & 0.2584 & 0.2586 & 0.2582 \\ 
 & (gain/loss) & -2.85\% \; & -87.31\% \; & -81.48\% \; & -66.22\% \; & +3.60\% & +3.27\% & +3.65\% \\
 \cline{2-9} 
 & fake-by-pop+reg & \textbf{0.0190} \,\; & 0.0181 * & 0.0193 * & \textbf{0.0063} * & \textbf{0.2602} & 0.1796 & \textbf{0.2772} \\ 
 & (gain/loss) & -3.55\% \; & -21.64\% \; & -20.57\% \; & -51.16\% \; & +18.16\% & +24.37\% & +17.40\% \\
 \hline
 \multirow{8}{*}{ml-10m} & reg & 0.0338 \,\; & 0.0215 * & 0.0213 * & 0.0198 * & 0.1625 & 0.1207 & 0.1654 \\  
 & (gain/loss) & -1.74\% \; & -15.01\% \; & -40.99\% \; & -42.93\% \; & -1.75\% & -1.38\% & -1.66\% \\
 \cline{2-9}
 & real+reg & \textbf{0.0381} * & 0.0033 \,\; & 0.0059 \,\; & 0.0029 \,\; & 0.1664 & 0.1599 & 0.1669 \\  
 & (gain/loss) & +26.15\% \; & -10.81\% \; & +25.53\% \; & $\ge$ +100\% & -4.03\% & -9.96\% & -3.63\% \\
 \cline{2-9}
 & fake+reg & 0.0337 \,\; & 0.0031 * & \textbf{0.0023} * & 0.0052 * & \textbf{0.1765} & \textbf{0.1941} & \textbf{0.1752} \\  
 & (gain/loss) & -1.74\% \; & -63.52\% \; & -70.12\% \; & -40.90\% \; & +2.31\% & +3.29\% & +2.15\% \\
 \cline{2-9}
 & fake-by-pop+reg & 0.0328 \,\; & \textbf{0.0019} * & \textbf{0.0023} * & \textbf{0.0004} * & 0.1684 & 0.1173 & 0.1718\\
 & (gain/loss) & -2.38\% \; & -89.90\% \; & -86.50\% \; & -97.66\% \; & +2.80\% & +9.72\% & +2.56\% \\
 \hline
\end{tabular}}
\caption{\textbf{Impact of Regularization on Recommended Lists}. Normalized Discounted Cumulative Gain (NDCG); Disparate Relevance ($\Delta \, \mathcal{R}$), Disparate Visibility ($\Delta \, \mathcal{V}$) and Disparate Exposure ($\Delta \, \mathcal{E}$) based on group contribution in the catalog; Coverage of the catalog ($\text{Cov}_{\text{tot}}$), of items from $\text{a}_{\min}$ ($\text{Cov}_{\text{a}_{\min}}$) and of items from  $\overline{\text{a}_{\min}}$ ($\text{Cov}_{\overline{\text{a}_{\min}}}$). We report the gain/loss of each regularized setting with respect to the non-regularized setting in Table \ref{tab:upsampling-1}.  Bold values refers to the best value across algorithms for a given dataset. (`*') indicates scores statistically different w.r.t. the non-regularized version ($p=0.05$).}
\label{tab:upsampling-2}
\end{table}

\subsubsection{Provider-level Walk-through Inspection of the Treatment (RQ3)}
Next, we analyze how our treatment affects the internal mechanisms of the user-item relevance learning step, and how these internal changes influence the recommended lists. To this end, we focus on a walk-through example of the problem and how our treatment addresses it. The goal is to understand where and how our treatment supports minority providers. 

\begin{figure}[!t]
\begin{subfigure}[t]{0.32\linewidth}
    \centering
    \includegraphics[width=1.0\linewidth]{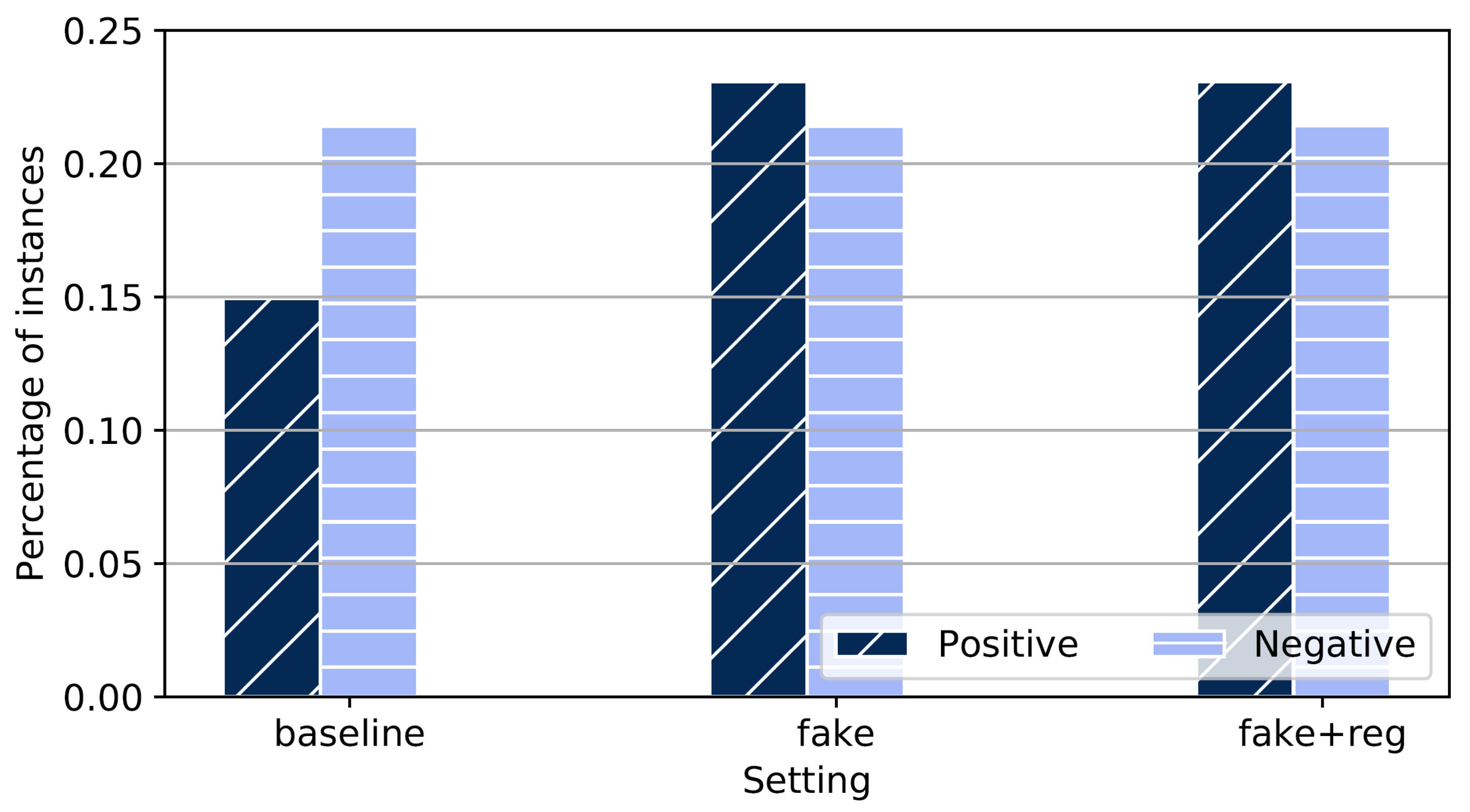}
    \caption{Minority instances}
    \label{fig:rq3-1}
\end{subfigure}
\begin{subfigure}[t]{0.32\linewidth}
    \centering
    \includegraphics[width=1.0\linewidth]{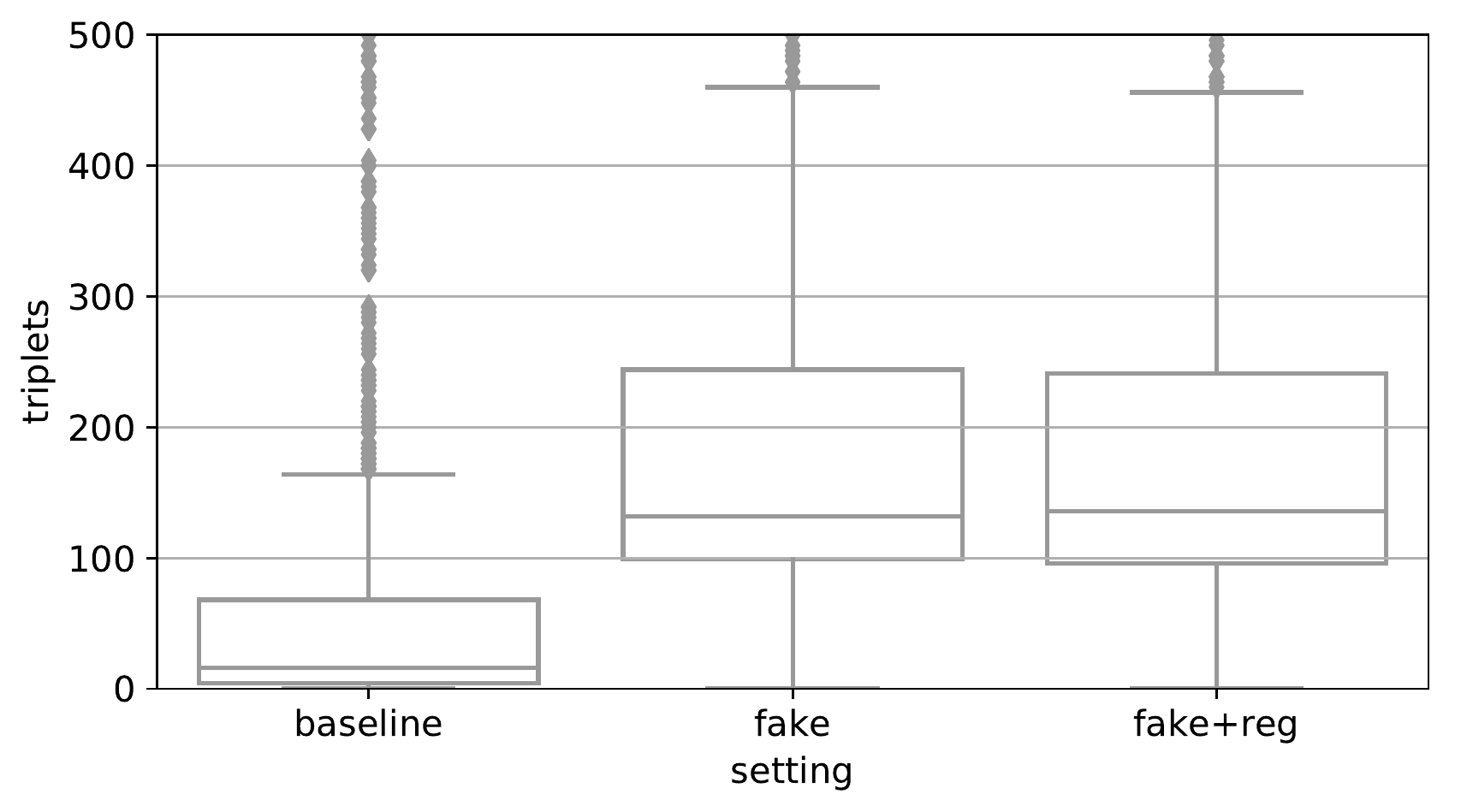}
    \caption{Provider triplets}
    \label{fig:rq3-2}
\end{subfigure}
\begin{subfigure}[t]{0.32\linewidth}
    \centering
    \includegraphics[width=1.0\linewidth]{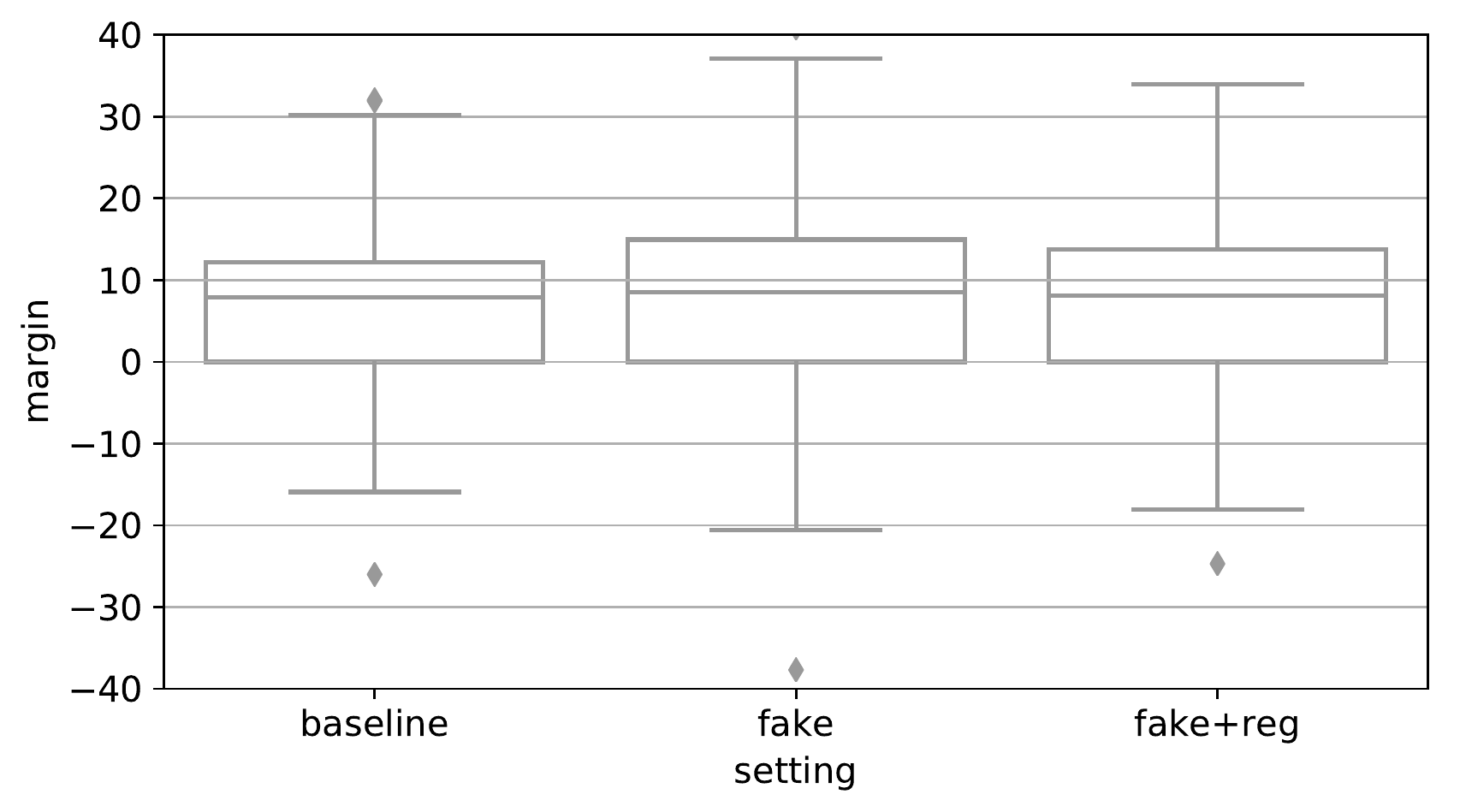}
    \caption{Provider margin}
    \label{fig:rq3-3}
\end{subfigure}
\begin{subfigure}[t]{0.33\linewidth}
    \centering
    \includegraphics[width=1.0\linewidth]{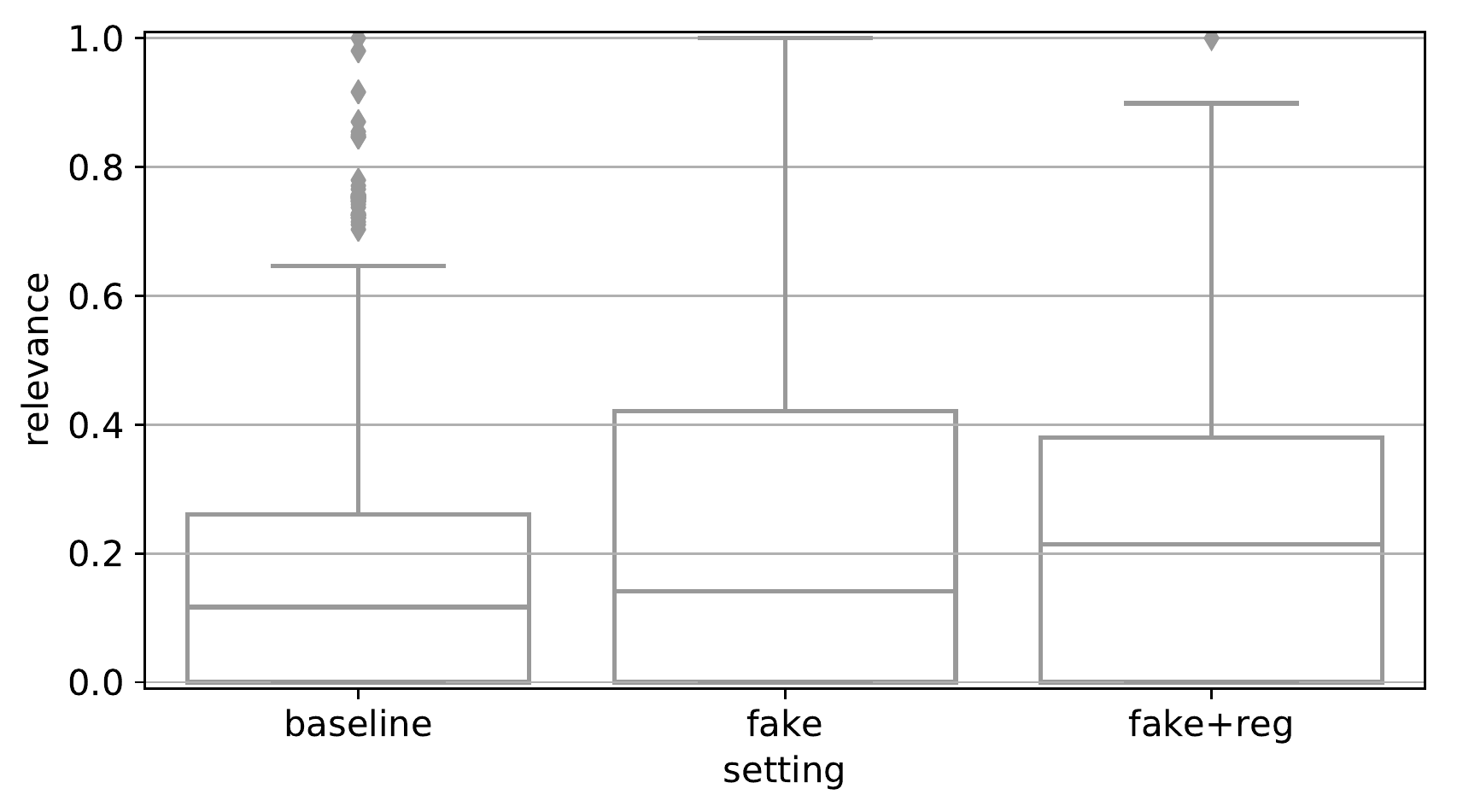}
    \caption{Provider relevance}
    \label{fig:rq3-4}
\end{subfigure}
\begin{subfigure}[t]{0.32\linewidth}
    \centering
    \includegraphics[width=1.0\linewidth]{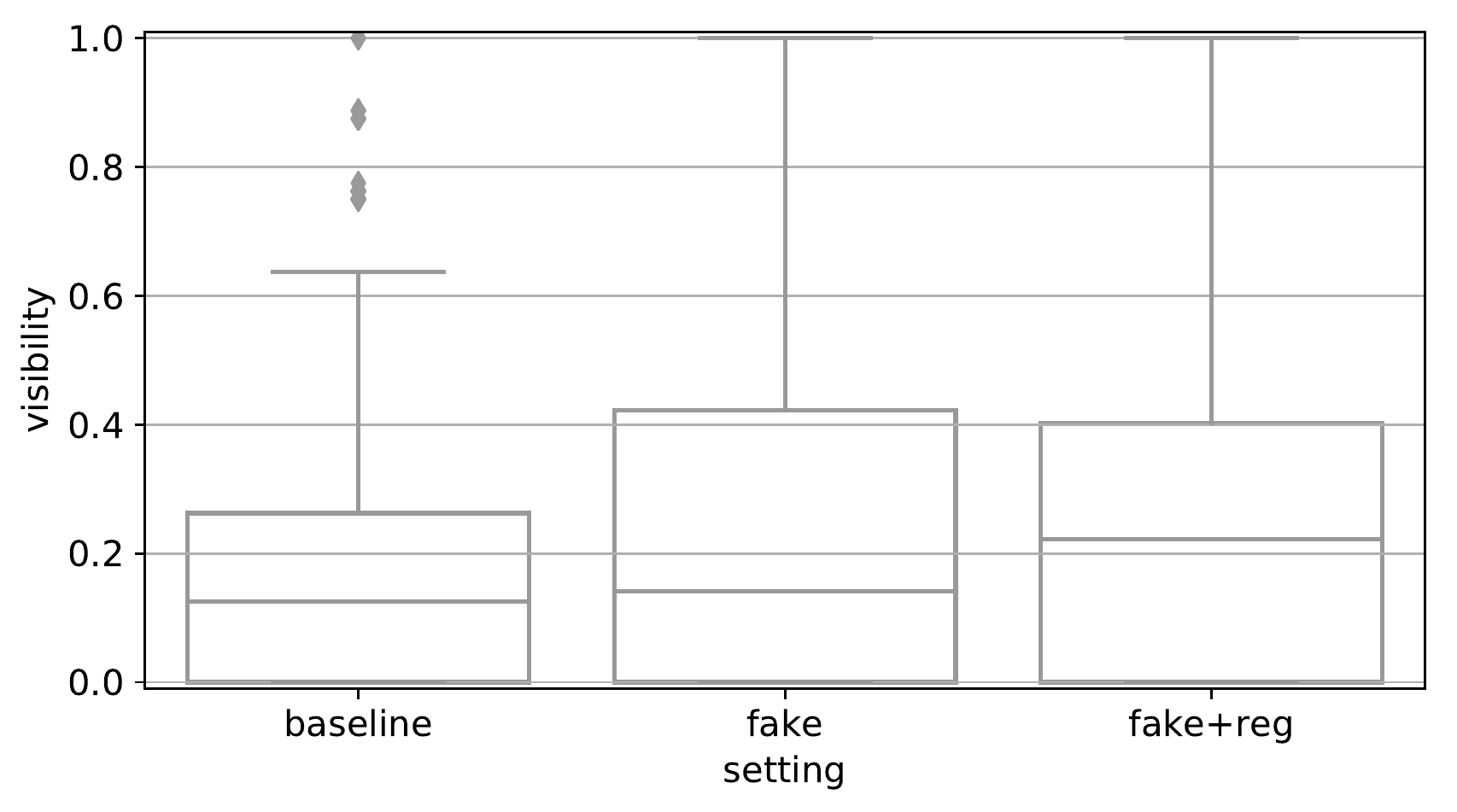}
    \caption{Provider visibility}
    \label{fig:rq3-5}
\end{subfigure}
\begin{subfigure}[t]{0.33\linewidth}
    \centering
    \includegraphics[width=1.0\linewidth]{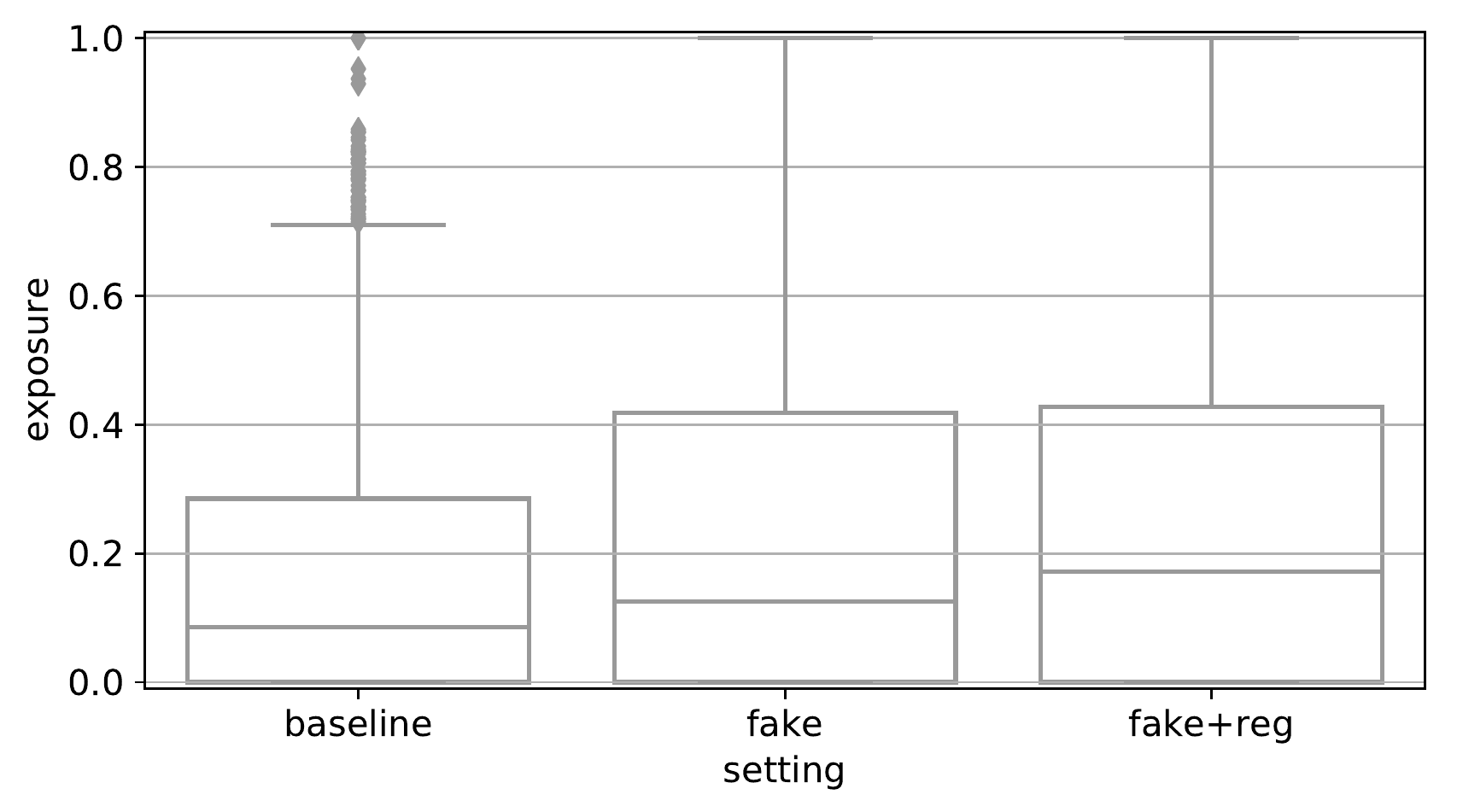}
    \caption{Provider exposure}
    \label{fig:rq3-6}
\end{subfigure}
\caption{\textbf{Walk-through Example}. Model properties concerning minority providers on \texttt{coco}, considering a baseline recommender and treatments with \texttt{fake} upsampling ($+0.09$ of minority data) and a regularization (with $\lambda=1e10^{-6}$). (a) number of triplets where the minority group is involved for the observed/unobserved item; (b) average number of triplets where a minority provider is involved for the observed item; (c) average margin between observed and unobserved items in a triplet, for triplets involving observed items of a minority provider; (d-f) average relevance, visibility, and exposure proportion assigned to items of the minority.}
\label{fig:rq3}
\end{figure}

To characterize our treatment, we consider the baseline recommender optimized on \texttt{coco} data. We are interested in showing how our treatment based on \texttt{fake} upsampling ($+0.09$ of minority data), followed by a regularization (with $\lambda=1e10^{-6}$), changes the internal and external properties shown by the baseline. Similar observations can be still applied to other settings. Figure \ref{fig:rq3-1} depicts the number of training triplets wherein an item delivered by a minority provider appears as a observed item (positive) or unobserved item (negative). Being under-represented in the interactions, items of minority-group providers appear less frequently as an observed item under the baseline setting (left-most pair of bars). It follows that the average number of triplets per provider, where a given minority provider is involved for the observed item is limited, as reported in Figure \ref{fig:rq3-2}  (left-most box plot). These imbalances strongly influence the ability of the pair-wise optimization of computing good margins between the observed and the unobserved item, when the former is delivered by a minority provider (Figure \ref{fig:rq3-3} - left-most box plot). With our upsampling, we introduce new user-item interactions involving minority providers, with more triplets for the minority group and a higher number of triplets per minority provider, on average (Figure \ref{fig:rq3-1} and \ref{fig:rq3-2} - two right-most box plots). This results in larger positive margins between observed and unobserved items for items of a minority provider (see Figure \ref{fig:rq3-3}, \texttt{fake} setting). Despite relying on the same up-sampled data, the regularized version further condenses the margins for observed items of minority providers around the average value (Figure \ref{fig:rq3-3}, \texttt{fake+reg} setting). This treatment fundamentally changes the relevance assigned to items for each minority provider and, by extension, their visibility and exposure, as highlighted in Figure \ref{fig:rq3-4}-\ref{fig:rq3-6}. 

\subsection{Comparing against Other Treatments (RQ4)}
We next compare our treatment against representative state-of-the-art alternatives to assess ($i$) how the considered treatments differently influence recommendations in terms of disparities, utility, and coverage, and ($ii$) whether the regularized relevance scores obtained through our treatment can lead to benefits for state-of-the-art mitigation procedures that operate in post-processing settings. Our goal in this section is to assess how far an in-processing strategy that reduces disparate relevance is from a post-processing strategy that directly controls exposure or visibility in rankings, in achieving good trade-offs. This experiment will provide evidence on the benefit of controlling the relevance distribution via upsampling and regularization. To this end, for each of the considered datasets, we decided to compare the recommendations generated after applying our \texttt{real+reg} treatment, which still use only real users' interactions, against those generated by the following three state-of-the-art mitigation procedures:
\begin{itemize}
\item \textbf{far} \cite{DBLP:conf/recsys/LiuGSBZ19} is a fairness criterion that combines a personalization-induced term and a fairness-induced term, with a parameter $\lambda$ controlling the trade-off between the two. The relevance score determined by the base recommender indicates the probability of a user being interested in an item. The fairness score promotes the items that belong to currently uncovered gender groups. We set up the size of ranked lists $k=10$, the trade-off parameter $\lambda=8.0$, and the desired percentage of minority items $p=C^{\text{a}_{\min}}$. 
\item \textbf{fa*ir} \cite{DBLP:conf/cikm/ZehlikeB0HMB17} is a fairness criterion based on maximizing utility while ensuring that the proportion of minority items in every prefix of the top-k ranking remains statistically above or indistinguishable from a given minimum, as long as there are enough minority items to achieve that minimum proportion. We set up the size of the considered ranking window $k=50$, the statistical significance parameter $\alpha=0.1$, and the parameters controlling the ranking prefixes $p=0.35$ for ml-10m and $p=0.50$ for coco. This will ensure that at least two and one ranking prefixes are used by the re-ranking, where two and one should be the minority items in a top-10 ranking for the datasets respectively, according to their $C^{\text{a}_{\min}}$ values.  
\item \textbf{fair-rec} \cite{DBLP:conf/www/PatroBGGC20} is a fairness criterion that, while maximizing utility and consumer fairness, aims to guarantee a uniform exposure distribution across providers. The first phase ensures user fairness among all the customers and tries to provide a minimum guarantee on exposure of the providers. Given that the first phase may not allocate exactly $k$ items to all the consumers, a second phase ensures this property while simultaneously maintaining provider fairness. We set up the size of the considered rankings $k=10$ and the fraction of share guarantee to be guaranteed to every provider $\alpha=0.5$.
\end{itemize}

These algorithms have been selected due to their different underlying approach and their ability to work with recommended lists. Each parameter in the corresponding algorithm has been reported after monitoring the trade-off between utility and disparate exposure. The first of the three algorithms has been re-implemented form scratch. The other ones were based on the original code provided by the authors in the paper\footnote{We have tailored our framework in order to produce the inputs needed by the algorithms.}.     

In order to answer to the first question, Table \ref{tab:baseline-1} provides ranking utility, disparity, and coverage scores for (i) our setting \texttt{real+reg} and (ii) the three other re-ranking algorithms fed with the relevance scores returned by the original recommender. It can be observed that our treatment \texttt{real+reg} has a good trade-off between utility and disparity. Specifically, an improvement of $19\%$ in NDCG on both datasets is introduced by our setup, when compared with the best NDCG among the other alternatives, i.e., \textbf{far}'s NDCG. In parallel, our treatment leads to the lowest disparate exposure, with a decrease of $97\%$ in coco and $48\%$ in ml-10m against \textbf{fa*ir} (the second best treatment in disparate exposure). It follows that our treatment reduces disparate exposure by moving up minority items which are also of interest for the consumers. In terms of coverage, \textbf{fair-rec} beats all the other treatments. However, in coco, such a higher coverage does not involve more items of the minority group. Indeed, while achieving an overall lower coverage, \texttt{real+reg} covers more items of the minority group with respect to \textbf{fair-rec}. Conversely, in ml-10m, the latter outperforms our treatment in minority group item coverage, but it leads to a NDCG lower of $26\%$ and to a disparate exposure higher of $78\%$, both being statistically significant. This allows us to draw the following observation:

\hlbox{Observation 6}{Upsampling minority items and regularizing relevance can often lead to higher recommendation utility and lower disparities, compared with considered treatments, regardless of the dataset. This benefit does not necessarily imply a higher coverage of minority items.}
 
\begin{table}[!t]
\resizebox{\textwidth}{!}{%
\begin{tabular}{cc|r|rrr|rrr}
\hline 
\textbf{Data} & \textbf{Type} & \textbf{NDCG} & \textbf{$\Delta \, \mathcal{R} $} & \textbf{$\Delta \, \mathcal{V}$} & \textbf{$\Delta \, \mathcal{E}$} & \textbf{$\text{Cov}_{\text{tot}}$} & \textbf{$\text{Cov}_{\text{a}_{\min}}$} & \textbf{$\text{Cov}_{\overline{\text{a}_{\min}}}$} \\
 \hline
\multirow{4}{*}{coco} & real+reg & \textbf{0.0178} * & \textbf{0.0134} * &  \textbf{0.0148} * & \textbf{0.0087} * & 0.2425 & \textbf{0.2728} & 0.2357 \\
 & far & 0.0150 * & 0.0810 * & 0.0812 * & 0.0763 * & 0.2206 & 0.1440 & 0.2368 \\
 & fa*ir & 0.0148 * & 0.0820 * & 0.0823 * & 0.0783 * & 0.2146 & 0.1403 & 0.2303 \\
 & fair-rec & 0.0139 * & 0.0800 * & 0.0800 * & 0.0776 * & \textbf{0.3039} & 0.1984 & \textbf{0.3262} \\
 \hline
\multirow{4}{*}{ml-10m} & real+reg & \textbf{0.0381} * & \textbf{0.0033} * & \textbf{0.0059} * & \textbf{0.0029} * & 0.1664 & 0.1599 & 0.1669 \\
 & far & 0.0318	* & 0.0370 \;\, & 0.0361 * & 0.0367 \;\, & 0.1787 & 0.1395 & 0.1814 \\
 & fa*ir & 0.0299 * & 0.0374 * & 0.0368 * & 0.0359	* & 0.1820 & 0.1463 & 0.1844 \\
 & fair-rec & 0.0300 * & 0.0331 * & 0.0314	* & 0.0275 * & \textbf{0.4335} & \textbf{0.3827} & \textbf{0.4369} \\
 \hline
\end{tabular}}
\caption{\textbf{Comparison against Other Treatments}. Normalized Discounted Cumulative Gain (NDCG); Disparate Relevance ($\Delta \, \mathcal{R}$), Disparate Visibility ($\Delta \, \mathcal{V}$) and Disparate Exposure ($\Delta \, \mathcal{E}$) based on minority contribution in the catalog; Coverage of the catalog ($\text{Cov}_{\text{tot}}$), of items from $\text{a}_{\min}$ ($\text{Cov}_{\text{a}_{\min}}$) and of items from  $\overline{\text{a}_{\min}}$ ($\text{Cov}_{\overline{\text{a}_{\min}}}$). Bold values indicate the best value across algorithms for each dataset. (`*') indicates scores statistically different with respect to the baseline ($p=0.05$).}
\label{tab:baseline-1}
\end{table}

\begin{table}[!t]
\resizebox{\textwidth}{!}{%
\begin{tabular}{cc|r|rrr|rrr}
\hline 
\textbf{Data} & \textbf{Type} & \textbf{NDCG} & \textbf{$\Delta \, \mathcal{R} $} & \textbf{$\Delta \, \mathcal{V}$} & \textbf{$\Delta \, \mathcal{E}$} & \textbf{$\text{Cov}_{\text{tot}}$} & \textbf{$\text{Cov}_{\text{a}_{\min}}$} & \textbf{$\text{Cov}_{\overline{\text{a}_{\min}}}$} \\
 \hline
\multirow{8}{*}{coco} & far & 0.0130 * & 0.0302 * & 0.0275 * & 0.0492 * & 0.2295 & 0.2565 & 0.2238 \\
&  (gain/loss)  & -13.35\% \; & -62.72\% \; & -66.13\% \;  & -37.16\% \; & +4.03\% & +78.12\% & -5.48\% \\
 \cline{2-9} 
& fa*ir & 0.0132 * & 0.0295	* & 0.0267	* & 0.0497	* & 0.2208 & 0.2476 & 0.2152 \\
&  (gain/loss)  & -10.81\% \; & -67.55\% \; & -36.52\% \;  & -64.02\% \;  & +2.89\% & +76.47\% & -6.55\% \\
 \cline{2-9} 
& fair-rec & 0.0132	* & 0.0230 * & 0.0264 *  & 0.0266 * & 0.3101 & 0.2958 & 0.3131 \\ 
&  (gain/loss)  & -5.04\% \; & -66.75\% \; & -71.25\% \; & 65.98\% \; & +2.04\% & +49.09\%  & -4.01\% \\
 \hline
\multirow{8}{*}{ml-10m} & far & \textbf{0.0351} * & \textbf{0.0005} *  & \textbf{0.0005} * & 0.0104 * & 0.1795 & 0.1684 & 0.1802 \\
&  (gain/loss)  & +10.38\% \; & -98.64\% \; & -98.61\% \; & -71.66\% \; & +0.44\% & +20.71\% & -0.66\% \\
 \cline{2-9} 
& fa*ir & 0.0321 * & 0.0003 * & 0.0005 * & 0.0118 * & 0.1821 & 0.1718 & 0.1828 \\
&  (gain/loss)  & +7.35\% \; &  -99.19\% \; & -98.64\% \; & -67.13\% \; & +0.05\% & +17.43\% & -0.87\% \\
 \cline{2-9} 
& fair-rec & 0.0325	* & 0.0043 * & 0.0031 * & \textbf{0.0039} * & \textbf{0.4342} & \textbf{0.4014} & \textbf{0.4364} \\
&  (gain/loss)  & +8.33\% \; & -87.00\% \; & -90.12\% \;  & -85.81\% \; & +0.16\% & +4.89\% & -0.11\% \\
 \hline
\end{tabular}}
\caption{\textbf{Benefits of Our Regularized Relevances to Other Treatments}. Normalized Discounted Cumulative Gain (NDCG); Disparate Relevance ($\Delta \, \mathcal{R}$), Disparate Visibility ($\Delta \, \mathcal{V}$) and Disparate Exposure ($\Delta \, \mathcal{E}$) based on group contribution in the catalog; Coverage of the catalog ($\text{Cov}_{\text{tot}}$), of items from $\text{a}_{\min}$ ($\text{Cov}_{\text{a}_{\min}}$) and of items from  $\overline{\text{a}_{\min}}$ ($\text{Cov}_{\overline{\text{a}_{\min}}}$). We report the gain/loss of each regularized setting with respect to the corresponding non-regularized setting in Table \ref{tab:baseline-1}. Bold values refers to the best value across algorithms for a given dataset. (`*') indicates scores statistically different with respect to the non-regularized version ($p=0.05$).}
\label{tab:baseline-2}
\end{table}

On the other hand, Table \ref{tab:baseline-2} allows us to answer to the second question, understanding whether the regularized relevance scores obtained through our treatment lead to benefits for state-of-the-art mitigation procedures that operate in post-processing settings. This table reports the utility, disparity, and coverage scores for the three state-of-the-art re-ranking algorithms fed with the relevance scores returned by our \texttt{real+reg} treatment, together with the relative improvement with respect to not using our non-regularized relevance scores. It can be observed that applying our treatment before re-ranking leads to a reduction of disparate exposure between $37\%$ and $86\%$. This improvement comes at the price of a negligible loss in accuracy in coco ($-5\%$ and $-13\%$), while the utility improves thanks to our regularized relevance scores in ml-10m (+$6\%$ and $+10\%)$. It follows that our treatment acts as a driver for improving the trade-off between effectiveness and disparities, highlighting the role of relevance scores in this context. In addition to this, all the settings show a higher overall coverage and coverage of minority items with respect to the non-regularized counterpart. Hence, we make the following observation:

\hlbox{Observation 7}{Relevances generated through minority-item upsampling and regularization enable the considered post-processing treatments to achieve a better trade-off between utility and disparity.}

Interestingly, comparing the results achieved with our treatment \texttt{real+reg} against those obtained with the other treatments under a regularized relevance setting, it can be observed that using the considered post-processing approaches fed with regularized relevances does not lead to substantial gains in utility and disparity trade-offs. 

Despite being related to disparate exposure, the fairness objective originally pursued by the considered countermeasures slightly differs from that targeted by this paper. Therefore, we also monitored the influence of our regularized relevances to those original fairness objectives. In \cite{DBLP:conf/recsys/LiuGSBZ19}, for \textbf{far}, Liu et al. monitored that each re-ranked list covers as many provider groups as possible. Under a \texttt{baseline} setting, the percentage of rankings covering both the providers' groups is $55\%$. This percentage increases to $98.4\%$ with \textbf{far} and to $99\%$ with \textbf{far} fed with the relevance scores returned by \texttt{real+reg}. Conversely, Zehlike et al. \cite{DBLP:conf/cikm/ZehlikeB0HMB17} used a ranked group fairness criterion that declares a ranking as unfair if the observed proportion of items from the minority group is far below the target one. Specifically, this criterion can be abstracted as comparing the number of protected items in every prefix of the ranking, with the expected number of protected items, if they were picked at random using Bernoulli trials. Under a \texttt{baseline} setting, the percentage of rankings that satisfy this criterion is $34.6\%$. This percentage increases to $78.7\%$ with \textbf{fa*ir} and to $80.2\%$ with \textbf{fa*ir} fed with the relevance scores returned by \texttt{real+reg}. Finally, given that the total exposure of the platform remains limited to $k*|U|$,  Patro et al. \cite{DBLP:conf/www/PatroBGGC20} aimed to guarantee that the items of each provider are recommended at least $(k*|U|) / |P|$ (i.e., this goal refers to the maximin marginal score value for the providers).  Under a \texttt{baseline} setting, the percentage of providers that satisfy this criterion is $23.3\%$. This percentage increases to $76.4\%$ with \textbf{fair-rec} and to $77.2\%$ with \textbf{fair-rec} fed with the relevance scores returned by \texttt{real+reg}. It follows that our approach not only leads to lower disparity, but also preserves the original objective of the post-processing algorithm and algorithm's utility. 

\subsection{Discussion}\label{sec:discussion}

Our experiments demonstrate that our intuitions were feasible for controlling the degree of share conveyed by relevance scores with respect to the contribution of the providers in the catalog. Our metric can be also optimized. 

Beyond our empirical work, we believe that \emph{our mapping approach} to associate providers' sensitive attributes to items sheds light on new perspectives of fairness in recommender systems. Many platforms include a range of items, whose mapping with the sensitive attributes of the providers is not as direct as in the case of items representing individuals. Existing approaches would move towards this direction and future fairness-aware recommendation approaches would require to embed this mapping to realistically shape real-world conditions. Indeed, this aspect will also drive the creation of new evaluation metrics and protocols, that allow to investigate algorithmic facets so far underexplored.   

Our study uncovered key connections among core components of optimization of recommendation models, while dealing with provider fairness. These results would promote even more the inspection of internal mechanisms in traditional strategies (e.g., pair-wise and point-wise), with a pro-active reaction to unfairness. Despite being relatively simple, our combination of upsampling and regularization provides fairness to target groups of providers, which could not be achieved individually by such components. Beyond being applied alone, our treatment can be envisioned as a pre-processing step for procedures that seek to have a fine-grained control of fairness, acting directly on recommended lists. In this case, our adjusted relevance scores can be used in post-processing fairness-aware procedures, possibly leading to a new space of optimization between fairness and recommendation utility. Our treatment is flexible enough to incorporate other strategies for controlling the share of relevance obtained through a recommendation algorithm, opening to interesting future work.

Since our study relied on a range of assumptions, we identified the main limitations of the approach presented in this paper, as listed below.
\begin{itemize}
\item The validity of the fairness notion we used is dependent on the integrity of the platform catalog, requiring to audit the catalog curation for sampling bias against direct discrimination (e.g., an educational platform that refuses to add courses provided by female instructors to its database).
\item Our empirical work dealt with scenarios with a very small minority, accounting for only $5\%-17\%$, depending on the dataset. There are many domains (or attributes) without this kind of minority; and this may lead to novel extensions and variants, starting from those suggested in this paper.
\item Experiments were based on a binary gender construct, with datasets providing only two genders, ``male'' and ``female''. Despite we had actually no chance of considering ``non-binary'' constructs, our formulation can be still applied to attributes with more than two genders. We remind readers to~\cite{DBLP:conf/chi/HamidiSB18} for consideration on the possible consequences of gender inference.
\item Grouping individuals in the COCO dataset relied on gender inference. However, this inference does not consider important elements, such as the intersectionality of gender with other sensitive attributes (e.g., geographic origin), the possibility of inferring non-binary gender labels, and how individuals self-identify themselves, since they capitalize on large historical databases. Being aware of this limitation, we did not provide any observation related to genders, and we used this dataset to assess the validity of our approach when a minority is present, regardless of the given gender being the minority. 
\item To better characterize our contributions, we focused on a matrix factorization approach optimized via pair-wise comparisons. Other variants could be tested with our framework as well, since our treatment does not rely on any specific peculiarity of the pair-wise optimization (we used it as it better aligns to top-$k$ recommendation problems).
\item Our approach does not have any mathematical guarantees on other notion of fairness in the recommended lists; however, we showed that it leads to a more balanced share of relevance and provides benefits to disparate impacts on visibility and exposure w.r.t. the contribution of the minority in the catalog. Further, our approach can be used as a pre-processing step for relevance scores, before using them with other treatments, e.g., \cite{DBLP:conf/sigir/BiegaGW18}).  
\end{itemize}

Despite these limitations, we believe that the intervention on the relevances we performed contribute to a better understanding of recommender systems.

\section{Related Work} \label{sec:related-work}
Our research is inspired by works on two areas that impact on recommender system research: (i) notions recently formalized in the context of fairness-aware rankings, and (ii) unfairness mitigation procedures on recommended lists.

\subsection{Provider Fairness Notions in Ranking and Recommendation}\label{sec:rel-met}
Fairness for groups traditionally requires that groups' exposure should be equally distributed between groups characterized by sensitive attributes (e.g., gender, race). Biega et al. \cite{DBLP:conf/sigir/BiegaGW18}, Singh and Joachims \cite{DBLP:conf/kdd/SinghJ18}, and Yadav et al. \cite{DBLP:journals/corr/abs-1911-08054} consider a notion of fairness based on equity. Despite working on provider groups, our work situates fairness in the context of recommender systems, allowing us to (i) account for situations where more providers lie behind an item and the same provider can appear more than once in a list, (ii) relate them with the objectives and formalism of recommendation metrics, and (iii) introduce a new experimentation on disparities among provider groups, starting from disparate relevance. Indeed, we control unfairness at an earlier stage, targeting the catalog contribution and not system-predicted relevance. 
Further, provider unfairness is traditionally mitigated by assuming to have access to true unbiased relevances. In practice, these relevances are estimated via machine learning, leading to a biased estimate of the relevance scores. Recommender systems are known to be biased from several perspectives (e.g., popularity, presentation, unfairness for users and providers). With this in mind, we control how relevance scores are distributed to groups. 

Comparing an outcome distribution (e.g., ranked lists) with a population distribution was explored by Yang and Stoyanovich \cite{DBLP:conf/ssdbm/YangS17} and Sapiezynski et al. \cite{DBLP:journals/corr/abs-1901-10437}. Differently from us, Sapiezynski et al. model uncertainty of group membership of a given individual, not dealing with contexts where more than one provider lies behind an item. Further, the outcome distribution is linked to a population distribution, assuming that the items the vendor chooses to show in the top-$k$ are a proportional representation of a subset of the catalog, sub-sampled via machine learning. This assumption may underestimate the real representation in the catalog, with respect to ours. Further, Yang and Stoyanovich compute the difference in the proportion of members of the protected group at top-$k$ and in the overall population. Compared to them, we internally control relevance according with contribution. Their formulations complement our ideas, as they drive fairness optimization at different levels.

Other fairness definitions in practice lead to enhanced fairness in exposure, for instance, by requiring equal proportions of individuals from different groups in ranking prefixes \cite{DBLP:conf/icalp/CelisSV18,DBLP:conf/cikm/ZehlikeB0HMB17,DBLP:conf/www/Zehlike020}. Mehrotra et al. \cite{DBLP:conf/cikm/MehrotraMBL018} achieved fairness through a re-ranking function, which balances accuracy and fairness by adding a personalized bonus to items of uncovered providers. Similarly, Burke et al. \cite{DBLP:conf/fat/BurkeSO18} define the concept of local fairness, and identified protected groups based on local conditions. In contrast to this, we study metrics that have a link between contribution, interactions, and relevance. The setup we study in this paper is very different, assuming that providers get relevance and, possibly, visibility and exposure, according to their contribution in the catalog. 

Furthermore, Patro et al. \cite{DBLP:conf/www/PatroBGGC20} account for uniform exposure over providers, while we deal with a relevance proportional to the providers' group contribution. Moreover, their definition assumes that items are not shareable, i.e., no item is allocated to multiple providers. Kamishima et al. \cite{DBLP:conf/fat/KamishimaAAS18} models fairness an as independence between the predicted rating values and sensitive values of the providers, not taking into account any measure against biased relevances (i.e., predicted rating), with respect to the contribution in the catalog. Beutel et al. \cite{DBLP:conf/kdd/BeutelCDQWWHZHC19} shapes fairness of providers in the context of pair-wise optimization, claiming fairness if the likelihood of an observed item being ranked above another relevant unclicked item is the same across both groups. Similarly, Narasimhan et al. \cite{DBLP:journals/corr/abs-1906-05330} propose a notion of pair-wise equal opportunity, requiring pairs to be equally-likely ranked correctly regardless of the group membership of both items in a pair. Compared with prior work, our approach aims to bind relevance and catalog contribution, for reducing disparate visibility and exposure. 

\subsection{Treatments for Provider Fairness} \label{sec:conn-treat}
There are relationships between our treatment and existing approaches, even though it should be trivial to consider that treatments fundamentally vary due to the different fairness notion they are driven by. 

Pre-processing for fairness in recommender systems has been considered in the context of consumer fairness. Rastegarpanah et al. \cite{DBLP:conf/wsdm/RastegarpanahGC19} proposed to add new fake users who provide ratings on existing items, to minimize the losses of all user's groups, computed as the mean squared estimation error over all known ratings in each group. Despite working on the provider side, our upsampling extends the interactions of the real users and items, and aims at adjusting interactions involving minority providers.  

In-processing regularization in recommender systems has traditionally focused on point-wise scenarios. Kamishima et al. \cite{DBLP:conf/fat/KamishimaAAS18} introduce a regularization requiring that the distance between the distribution of predicted ratings for items belonging to two different groups is as small as possible. However, this way of optimizing does not indicate much about the resulting recommended lists that users actually see, with respect to the pair-wise optimization we leveraged, and do not take into account to what degree ratings of different groups are proportional to the group contribution in the catalog. 

Beutel et al. \cite{DBLP:conf/kdd/BeutelCDQWWHZHC19} targeted provider fairness optimization, under a pair-wise optimization scenario, similarly to us. However, while the pair-wise comparisons are at the basis of their fairness definition, our treatment is just tested under a pair-wise optimization scenario and does not leverage any peculiarity of this scenario. Further, while both have been tested on binary attributes, we generalize to capture a wider variety of groups, and generalize to contexts where items are associated to more than one provider. Their training methodology is also very different. The fixed regularization term they added to the loss function is based on a \emph{correlation} between the residual estimate and the group membership. These conceptual and operative differences lead us to investigate clearly different under-explored facets. Furthermore, compared to our work, they are driven by a different fairness objective. It would be interesting to see how they can be integrated, taking the benefits of both the notions, but this requires non-trivial extensions left as a future work. 

Finally, other fairness-aware approaches, whose notions of fairness were presented in the previous section, are operationalized in quite different ways. Biega et al. \cite{DBLP:conf/sigir/BiegaGW18} solve an integer linear program. Patro et al. \cite{DBLP:conf/www/PatroBGGC20} implement a Greedy-round-robin strategy. Similarly to us, Zehlike and Castillo \cite{DBLP:conf/www/Zehlike020} use stochastic gradient descent, but they operationalize it in a list-wise manner.

\section{Conclusions} \label{sec:conclusions}
In this paper, we assessed the extent to which a recommender system emphasizes disparities from three different perspectives. The first one monitors the difference of predicted relevance between providers' groups, according with their representation in the catalog. The other two operate on the final outcomes of the recommender system, by monitoring differences in visibility and exposure across groups, with the same proportional setting. To reduce the emerged disparities, we proposed a treatment that combines an upsampling of interactions from the minority group and a regularization on the share of relevance across provider groups, throughout the optimization process.

Our experimental study analyzes relevance scores and recommended lists generated by fifteen synthetic datasets that simulate specific situations of imbalance in the catalog and in the interactions, and two real-world datasets that represent existent conditions in modern platforms. Our first exploratory results highlight that the discrepancy between the relevance given to provider groups by recommendation models and their contribution in the catalog is not negligible. This effect results in less than expected visibility and exposure for the minority group. With our treatment, it has been possible to reduce disparities in relevance, visibility and exposure, without sacrificing recommendation utility. Incorporating our treatment allows to act indirectly on the output of the recommendation model and is a viable strategy to mitigate distortions at an earlier step. Our treatment has been proved to be useful also for post-processing fairness procedures in order to achieve lower disparities.

Future  work  will  embrace  all the  insights  provided  in  this  paper  to  further explore  the  connection  between  relevance, visibility, and exposure. Moreover, we plan to design mitigation methods that look at  provider  fairness  promotion  as  a  temporal  process.  The  improvement  in provider fairness might not be large immediately, and we believe that repeating our treatment over time will lead to more and more fair recommendations. This would better fit with real-world situations and platforms. We will also investigate the relationships between the recommendations returned by the algorithm and the tendency of each user to prefer items from different groups of providers. Finally, it is our goal to devise other treatments that link internal parameters to ranking metrics.

\begin{acknowledgements}
Mirko Marras acknowledges Sardinia Regional Government for the financial support of his PhD scholarship (P.O.R. Sardegna F.S.E.  Operational Programme of the Autonomous Region of Sardinia, European Social Fund 2014-2020   - Axis III Education and Training, Thematic Goal 10, Priority of Investment 10ii, Specific Goal 10.5). Ludovico Boratto acknowledges Ag\`encia per a la Competivitat de l'Empresa, ACCI\'O, for their support under project "Privacy-preserving, Fair and Explainable Artificial Intelligence (PrEFair)". 
\end{acknowledgements}

\appendix

\section{Synthetic Datasets Creation} \label{sec:syn-data}
In order to simulate catalog and interaction imbalance through synthetic datasets, we use two stochastic block models \cite{DBLP:conf/nips/YaoH17}. We create a \texttt{catalog block model} to determine the probability that an item is offered by a provider in a particular group. Non-uniformity in this block model will lead to catalog imbalances. We then arrange an \texttt{observation block model}, determining the probability that a user observes an item from a given provider's group, simulating an implicit feedback scenario. The group ratios may be non-uniform, leading to interaction imbalance. Formally, let vector $L \in [0, 1]^{|A|}$, with $A = \{a_{1}, a_{2}\}$, be the block-model parameters for catalog probability. For an item $i$, the probability of assigning it to a provider with $a_i$ is $L(a_i)$. Moreover, given a user $u$, let $O \in [0, 1]^{|A|}$ be such that the probability of observing an item $i$ with a provider having $a_i$ is $O(a_i)$. Specifically, based on groups in $A$, we consider five catalog block models $L_x = [x, 1-x]$ and five observation block models $O_y = [y, 1-y]$, with $x,y \in V=\{0.1, 0.2, 0.3, 0.4, 0.5\}$. To replicate our target recommendation context, where interaction imbalance is assumed to be equal or higher than catalog imbalance, our study will consider $15$ setups $(L_x, O_y)$, with $x,y \in V$ and $x \ge y$. Hence, our exploration will cover both situations with a small minority and where the groups are more balanced. Specifically, providers in $a_1$ are identified as the minority group, i.e., $\text{a}_{\min}$.  

For each setup $(L_x, O_y)$, we selected a catalog block model and an observation block model, (i) generating $n=30,000$ users and $m=3,000$ items, (ii) assigning catalog representations based on $L_x$, and (iii) sampling $o=1,200,000$ implicit interactions, according with $O_y$. This step means that we randomly sampled a user $u$, then we selected the provider group $a \in A$ of the item in that pair according to $O_y$, and we sampled an item $i$ from the selected group. To limit anomalous results and distorted recommendation outputs, for each provider group, our specific procedure samples the item $i \in I_a$ simulating a scenario where items in the same provider group have a different probability of being selected. To this end, we used an exponential distribution $\mathcal{X}$ with scale $\omega$ for the distribution function $\phi$. The parameter $\omega$ determines the scale of the exponential distribution, with $\phi(\omega) = \mathcal{E} (0, (\frac{|I_a|}{\omega})^2)$. Given the list $L_{I_a}$ of items in $I_a$ and the distribution $\phi(\omega)$, the index of the sample item $i$ in $L_{I_a}$ is represented by the absolute rounded value of the random variable $\phi(\omega)$. Decreasing $\omega$ means that we make the selection more uniform across items. Our exploratory study was carried out with $\omega=450$, with the aim of reflecting realistic popularity tails. At the end, we obtained $15$ synthetic datasets with different representations of the minority group in the catalog and the interactions.

\bibliographystyle{spmpsci}     
\bibliography{sample-base}  

\begin{thebibliography}{10}
\providecommand{\url}[1]{{#1}}
\providecommand{\urlprefix}{URL }
\expandafter\ifx\csname urlstyle\endcsname\relax
  \providecommand{\doi}[1]{DOI~\discretionary{}{}{}#1}\else
  \providecommand{\doi}{DOI~\discretionary{}{}{}\begingroup
  \urlstyle{rm}\Url}\fi

\bibitem{DBLP:conf/kdd/BeutelCDQWWHZHC19}
Beutel, A., Chen, J., Doshi, T., Qian, H., Wei, L., Wu, Y., Heldt, L., Zhao,
  Z., Hong, L., Chi, E.H., Goodrow, C.: Fairness in recommendation ranking
  through pairwise comparisons.
\newblock In: A.~Teredesai, V.~Kumar, Y.~Li, R.~Rosales, E.~Terzi, G.~Karypis
  (eds.) Proceedings of the 25th {ACM} {SIGKDD} International Conference on
  Knowledge Discovery {\&} Data Mining, {KDD} 2019, Anchorage, AK, USA, August
  4-8, 2019, pp. 2212--2220. {ACM} (2019).
\newblock \doi{10.1145/3292500.3330745}.
\newblock \urlprefix\url{https://doi.org/10.1145/3292500.3330745}

\bibitem{DBLP:journals/corr/abs-2003-11650}
Biega, A.J., Diaz, F., Ekstrand, M.D., Kohlmeier, S.: Overview of the {TREC}
  2019 fair ranking track.
\newblock CoRR \textbf{abs/2003.11650} (2020).
\newblock \urlprefix\url{https://arxiv.org/abs/2003.11650}

\bibitem{DBLP:conf/sigir/BiegaGW18}
Biega, A.J., Gummadi, K.P., Weikum, G.: Equity of attention: Amortizing
  individual fairness in rankings.
\newblock In: K.~Collins{-}Thompson, Q.~Mei, B.D. Davison, Y.~Liu, E.~Yilmaz
  (eds.) The 41st International {ACM} {SIGIR} Conference on Research {\&}
  Development in Information Retrieval, {SIGIR} 2018, Ann Arbor, MI, USA, July
  08-12, 2018, pp. 405--414. {ACM} (2018).
\newblock \doi{10.1145/3209978.3210063}.
\newblock \urlprefix\url{https://doi.org/10.1145/3209978.3210063}

\bibitem{DBLP:journals/ipm/BorattoFM21}
Boratto, L., Fenu, G., Marras, M.: Connecting user and item perspectives in
  popularity debiasing for collaborative recommendation.
\newblock Inf. Process. Manag. \textbf{58}(1), 102387 (2021).
\newblock \doi{10.1016/j.ipm.2020.102387}.
\newblock \urlprefix\url{https://doi.org/10.1016/j.ipm.2020.102387}

\bibitem{DBLP:journals/corr/Burke17aa}
Burke, R.: Multisided fairness for recommendation.
\newblock CoRR \textbf{abs/1707.00093} (2017).
\newblock \urlprefix\url{http://arxiv.org/abs/1707.00093}

\bibitem{DBLP:conf/fat/BurkeSO18}
Burke, R., Sonboli, N., Ordonez{-}Gauger, A.: Balanced neighborhoods for
  multi-sided fairness in recommendation.
\newblock In: S.A. Friedler, C.~Wilson (eds.) Conference on Fairness,
  Accountability and Transparency, {FAT} 2018, 23-24 February 2018, New York,
  NY, {USA}, \emph{Proceedings of Machine Learning Research}, vol.~81, pp.
  202--214. {PMLR} (2018).
\newblock \urlprefix\url{http://proceedings.mlr.press/v81/burke18a.html}

\bibitem{DBLP:journals/umuai/CamposDC14}
Campos, P.G., D{\'{\i}}ez, F., Cantador, I.: Time-aware recommender systems: a
  comprehensive survey and analysis of existing evaluation protocols.
\newblock User Model. User-Adapt. Interact. \textbf{24}(1-2), 67--119 (2014).
\newblock \doi{10.1007/s11257-012-9136-x}.
\newblock \urlprefix\url{https://doi.org/10.1007/s11257-012-9136-x}

\bibitem{DBLP:conf/icalp/CelisSV18}
Celis, L.E., Straszak, D., Vishnoi, N.K.: Ranking with fairness constraints.
\newblock In: I.~Chatzigiannakis, C.~Kaklamanis, D.~Marx, D.~Sannella (eds.)
  45th International Colloquium on Automata, Languages, and Programming,
  {ICALP} 2018, July 9-13, 2018, Prague, Czech Republic, \emph{LIPIcs}, vol.
  107, pp. 28:1--28:15. Schloss Dagstuhl - Leibniz-Zentrum f{\"{u}}r Informatik
  (2018).
\newblock \doi{10.4230/LIPIcs.ICALP.2018.28}.
\newblock \urlprefix\url{https://doi.org/10.4230/LIPIcs.ICALP.2018.28}

\bibitem{chen2017attentive}
Chen, J., Zhang, H., He, X., Nie, L., Liu, W., Chua, T.S.: Attentive
  collaborative filtering: Multimedia recommendation with item-and
  component-level attention.
\newblock In: Proceedings of the 40th International ACM SIGIR conference on
  Research and Development in Information Retrieval, pp. 335--344. ACM (2017)

\bibitem{DBLP:conf/chi/ChenMHW18}
Chen, L., Ma, R., Hann{\'{a}}k, A., Wilson, C.: Investigating the impact of
  gender on rank in resume search engines.
\newblock In: R.L. Mandryk, M.~Hancock, M.~Perry, A.L. Cox (eds.) Proceedings
  of the 2018 {CHI} Conference on Human Factors in Computing Systems, {CHI}
  2018, Montreal, QC, Canada, April 21-26, 2018, p. 651. {ACM} (2018).
\newblock \doi{10.1145/3173574.3174225}.
\newblock \urlprefix\url{https://doi.org/10.1145/3173574.3174225}

\bibitem{DBLP:conf/worldcist/DessiFMR18}
Dess{\`{\i}}, D., Fenu, G., Marras, M., Recupero, D.R.: {COCO:}
  semantic-enriched collection of online courses at scale with experimental use
  cases.
\newblock In: {\'{A}}.~Rocha, H.~Adeli, L.P. Reis, S.~Costanzo (eds.) Trends
  and Advances in Information Systems and Technologies - Volume 2
  [WorldCIST'18, Naples, Italy, March 27-29, 2018], \emph{Advances in
  Intelligent Systems and Computing}, vol. 746, pp. 1386--1396. Springer
  (2018).
\newblock \doi{10.1007/978-3-319-77712-2\_133}.
\newblock \urlprefix\url{https://doi.org/10.1007/978-3-319-77712-2\_133}

\bibitem{DBLP:conf/innovations/DworkHPRZ12}
Dwork, C., Hardt, M., Pitassi, T., Reingold, O., Zemel, R.S.: Fairness through
  awareness.
\newblock In: S.~Goldwasser (ed.) Innovations in Theoretical Computer Science
  2012, Cambridge, MA, USA, January 8-10, 2012, pp. 214--226. {ACM} (2012).
\newblock \doi{10.1145/2090236.2090255}.
\newblock \urlprefix\url{https://doi.org/10.1145/2090236.2090255}

\bibitem{DBLP:conf/chi/HamidiSB18}
Hamidi, F., Scheuerman, M.K., Branham, S.M.: Gender recognition or gender
  reductionism?: The social implications of embedded gender recognition
  systems.
\newblock In: R.L. Mandryk, M.~Hancock, M.~Perry, A.L. Cox (eds.) Proceedings
  of the 2018 {CHI} Conference on Human Factors in Computing Systems, {CHI}
  2018, Montreal, QC, Canada, April 21-26, 2018, p.~8. {ACM} (2018).
\newblock \doi{10.1145/3173574.3173582}.
\newblock \urlprefix\url{https://doi.org/10.1145/3173574.3173582}

\bibitem{DBLP:journals/tiis/HarperK16}
Harper, F.M., Konstan, J.A.: The movielens datasets: History and context.
\newblock {ACM} Trans. Interact. Intell. Syst. \textbf{5}(4), 19:1--19:19
  (2016).
\newblock \doi{10.1145/2827872}.
\newblock \urlprefix\url{https://doi.org/10.1145/2827872}

\bibitem{DBLP:journals/tmis/JannachJ19}
Jannach, D., Jugovac, M.: Measuring the business value of recommender systems.
\newblock {ACM} Trans. Management Inf. Syst. \textbf{10}(4), 16:1--16:23
  (2019).
\newblock \doi{10.1145/3370082}.
\newblock \urlprefix\url{https://doi.org/10.1145/3370082}

\bibitem{DBLP:journals/tois/JarvelinK02}
J{\"{a}}rvelin, K., Kek{\"{a}}l{\"{a}}inen, J.: Cumulated gain-based evaluation
  of {IR} techniques.
\newblock {ACM} Trans. Inf. Syst. \textbf{20}(4), 422--446 (2002).
\newblock \doi{10.1145/582415.582418}.
\newblock \urlprefix\url{http://doi.acm.org/10.1145/582415.582418}

\bibitem{DBLP:journals/tiis/KaminskasB17}
Kaminskas, M., Bridge, D.: Diversity, serendipity, novelty, and coverage: {A}
  survey and empirical analysis of beyond-accuracy objectives in recommender
  systems.
\newblock {ACM} Trans. Interact. Intell. Syst. \textbf{7}(1), 2:1--2:42 (2017).
\newblock \doi{10.1145/2926720}.
\newblock \urlprefix\url{https://doi.org/10.1145/2926720}

\bibitem{DBLP:conf/fat/KamishimaAAS18}
Kamishima, T., Akaho, S., Asoh, H., Sakuma, J.: Recommendation independence.
\newblock In: S.A. Friedler, C.~Wilson (eds.) Conference on Fairness,
  Accountability and Transparency, {FAT} 2018, 23-24 February 2018, New York,
  NY, {USA}, \emph{Proceedings of Machine Learning Research}, vol.~81, pp.
  187--201. {PMLR} (2018).
\newblock \urlprefix\url{http://proceedings.mlr.press/v81/kamishima18a.html}

\bibitem{koren2009matrix}
Koren, Y., Bell, R., Volinsky, C.: Matrix factorization techniques for
  recommender systems.
\newblock Computer (8), 30--37 (2009)

\bibitem{DBLP:journals/pvldb/LahotiGW19}
Lahoti, P., Gummadi, K.P., Weikum, G.: Operationalizing individual fairness
  with pairwise fair representations.
\newblock Proc. {VLDB} Endow. \textbf{13}(4), 506--518 (2019).
\newblock \urlprefix\url{http://www.vldb.org/pvldb/vol13/p506-lahoti.pdf}

\bibitem{DBLP:journals/ajiips/LiuSZGX19}
Liu, B., Su, Y., Zha, D., Gao, N., Xiang, J.: Carec: Content-aware
  point-of-interest recommendation via adaptive bayesian personalized ranking.
\newblock Aust. J. Intell. Inf. Process. Syst. \textbf{15}(3), 61--68 (2019).
\newblock \urlprefix\url{http://ajiips.com.au/papers/V15.4/v15n4\_65-72.pdf}

\bibitem{DBLP:conf/recsys/LiuGSBZ19}
Liu, W., Guo, J., Sonboli, N., Burke, R., Zhang, S.: Personalized
  fairness-aware re-ranking for microlending.
\newblock In: T.~Bogers, A.~Said, P.~Brusilovsky, D.~Tikk (eds.) Proceedings of
  the 13th {ACM} Conference on Recommender Systems, RecSys 2019, Copenhagen,
  Denmark, September 16-20, 2019, pp. 467--471. {ACM} (2019).
\newblock \doi{10.1145/3298689.3347016}.
\newblock \urlprefix\url{https://doi.org/10.1145/3298689.3347016}

\bibitem{DBLP:conf/recsys/MansouryMBP19}
Mansoury, M., Mobasher, B., Burke, R., Pechenizkiy, M.: Bias disparity in
  collaborative recommendation: Algorithmic evaluation and comparison.
\newblock In: R.~Burke, H.~Abdollahpouri, E.C. Malthouse, K.P. Thai, Y.~Zhang
  (eds.) Proceedings of the Workshop on Recommendation in Multi-stakeholder
  Environments co-located with the 13th {ACM} Conference on Recommender Systems
  (RecSys 2019), Copenhagen, Denmark, September 20, 2019, \emph{{CEUR} Workshop
  Proceedings}, vol. 2440. CEUR-WS.org (2019).
\newblock \urlprefix\url{http://ceur-ws.org/Vol-2440/paper6.pdf}

\bibitem{DBLP:conf/cikm/MehrotraMBL018}
Mehrotra, R., McInerney, J., Bouchard, H., Lalmas, M., Diaz, F.: Towards a fair
  marketplace: Counterfactual evaluation of the trade-off between relevance,
  fairness {\&} satisfaction in recommendation systems.
\newblock In: A.~Cuzzocrea, J.~Allan, N.W. Paton, D.~Srivastava, R.~Agrawal,
  A.Z. Broder, M.J. Zaki, K.S. Candan, A.~Labrinidis, A.~Schuster, H.~Wang
  (eds.) Proceedings of the 27th {ACM} International Conference on Information
  and Knowledge Management, {CIKM} 2018, Torino, Italy, October 22-26, 2018,
  pp. 2243--2251. {ACM} (2018).
\newblock \doi{10.1145/3269206.3272027}.
\newblock \urlprefix\url{https://doi.org/10.1145/3269206.3272027}

\bibitem{DBLP:journals/corr/abs-1906-05330}
Narasimhan, H., Cotter, A., Gupta, M.R., Wang, S.: Pairwise fairness for
  ranking and regression.
\newblock CoRR \textbf{abs/1906.05330} (2019).
\newblock \urlprefix\url{http://arxiv.org/abs/1906.05330}

\bibitem{DBLP:conf/www/PatroBGGC20}
Patro, G.K., Biswas, A., Ganguly, N., Gummadi, K.P., Chakraborty, A.: Fairrec:
  Two-sided fairness for personalized recommendations in two-sided platforms.
\newblock In: Y.~Huang, I.~King, T.~Liu, M.~van Steen (eds.) {WWW} '20: The Web
  Conference 2020, Taipei, Taiwan, April 20-24, 2020, pp. 1194--1204. {ACM} /
  {IW3C2} (2020).
\newblock \doi{10.1145/3366423.3380196}.
\newblock \urlprefix\url{https://doi.org/10.1145/3366423.3380196}

\bibitem{DBLP:conf/wsdm/RastegarpanahGC19}
Rastegarpanah, B., Gummadi, K.P., Crovella, M.: Fighting fire with fire: Using
  antidote data to improve polarization and fairness of recommender systems.
\newblock In: J.S. Culpepper, A.~Moffat, P.N. Bennett, K.~Lerman (eds.)
  Proceedings of the Twelfth {ACM} International Conference on Web Search and
  Data Mining, {WSDM} 2019, Melbourne, VIC, Australia, February 11-15, 2019,
  pp. 231--239. {ACM} (2019).
\newblock \doi{10.1145/3289600.3291002}.
\newblock \urlprefix\url{https://doi.org/10.1145/3289600.3291002}

\bibitem{rendle2012bpr}
Rendle, S., Freudenthaler, C., Gantner, Z., Schmidt-Thieme, L.: Bpr: Bayesian
  personalized ranking from implicit feedback.
\newblock arXiv preprint arXiv:1205.2618  (2012)

\bibitem{DBLP:reference/sp/RicciRS15}
Ricci, F., Rokach, L., Shapira, B.: Recommender systems: Introduction and
  challenges.
\newblock In: F.~Ricci, L.~Rokach, B.~Shapira (eds.) Recommender Systems
  Handbook, pp. 1--34. Springer (2015).
\newblock \doi{10.1007/978-1-4899-7637-6\_1}.
\newblock \urlprefix\url{https://doi.org/10.1007/978-1-4899-7637-6\_1}

\bibitem{SanchezB20}
S\'{a}nchez, P., Bellog\'{i}n, A.: Applying reranking strategies to route
  recommendation using sequence-aware evaluation.
\newblock User Model. User-Adapt. Interact.  (2020)

\bibitem{DBLP:journals/corr/abs-1901-10437}
Sapiezynski, P., Zeng, W., Robertson, R.E., Mislove, A., Wilson, C.:
  Quantifying the impact of user attention on fair group representation in
  ranked lists.
\newblock CoRR \textbf{abs/1901.10437} (2019).
\newblock \urlprefix\url{http://arxiv.org/abs/1901.10437}

\bibitem{DBLP:conf/kdd/SinghJ18}
Singh, A., Joachims, T.: Fairness of exposure in rankings.
\newblock In: Y.~Guo, F.~Farooq (eds.) Proceedings of the 24th {ACM} {SIGKDD}
  International Conference on Knowledge Discovery {\&} Data Mining, {KDD} 2018,
  London, UK, August 19-23, 2018, pp. 2219--2228. {ACM} (2018).
\newblock \doi{10.1145/3219819.3220088}.
\newblock \urlprefix\url{https://doi.org/10.1145/3219819.3220088}

\bibitem{DBLP:book/Walster1973new}
Walster, E., Berscheid, E., Walster, G.W.: New directions in equity research.
\newblock Journal of personality and social psychology \textbf{25}(2), 151
  (1973)

\bibitem{xiao2017attentional}
Xiao, J., Ye, H., He, X., Zhang, H., Wu, F., Chua, T.S.: Attentional
  factorization machines: Learning the weight of feature interactions via
  attention networks.
\newblock arXiv preprint arXiv:1708.04617  (2017)

\bibitem{xue2017deep}
Xue, H.J., Dai, X., Zhang, J., Huang, S., Chen, J.: Deep matrix factorization
  models for recommender systems.
\newblock In: IJCAI, pp. 3203--3209 (2017)

\bibitem{DBLP:journals/corr/abs-1911-08054}
Yadav, H., Du, Z., Joachims, T.: Fair learning-to-rank from implicit feedback.
\newblock CoRR \textbf{abs/1911.08054} (2019).
\newblock \urlprefix\url{http://arxiv.org/abs/1911.08054}

\bibitem{DBLP:conf/ssdbm/YangS17}
Yang, K., Stoyanovich, J.: Measuring fairness in ranked outputs.
\newblock In: Proceedings of the 29th International Conference on Scientific
  and Statistical Database Management, Chicago, IL, USA, June 27-29, 2017, pp.
  22:1--22:6. {ACM} (2017).
\newblock \doi{10.1145/3085504.3085526}.
\newblock \urlprefix\url{https://doi.org/10.1145/3085504.3085526}

\bibitem{DBLP:conf/nips/YaoH17}
Yao, S., Huang, B.: Beyond parity: Fairness objectives for collaborative
  filtering.
\newblock In: I.~Guyon, U.~von Luxburg, S.~Bengio, H.M. Wallach, R.~Fergus,
  S.V.N. Vishwanathan, R.~Garnett (eds.) Advances in Neural Information
  Processing Systems 30: Annual Conference on Neural Information Processing
  Systems 2017, 4-9 December 2017, Long Beach, CA, {USA}, pp. 2921--2930
  (2017).
\newblock
  \urlprefix\url{http://papers.nips.cc/paper/6885-beyond-parity-fairness-objectives-for-collaborative-filtering}

\bibitem{DBLP:conf/cikm/ZehlikeB0HMB17}
Zehlike, M., Bonchi, F., Castillo, C., Hajian, S., Megahed, M., Baeza{-}Yates,
  R.: Fa*ir: {A} fair top-k ranking algorithm.
\newblock In: E.~Lim, M.~Winslett, M.~Sanderson, A.W. Fu, J.~Sun, J.S.
  Culpepper, E.~Lo, J.C. Ho, D.~Donato, R.~Agrawal, Y.~Zheng, C.~Castillo,
  A.~Sun, V.S. Tseng, C.~Li (eds.) Proceedings of the 2017 {ACM} on Conference
  on Information and Knowledge Management, {CIKM} 2017, Singapore, November 06
  - 10, 2017, pp. 1569--1578. {ACM} (2017).
\newblock \doi{10.1145/3132847.3132938}.
\newblock \urlprefix\url{https://doi.org/10.1145/3132847.3132938}

\bibitem{DBLP:conf/www/Zehlike020}
Zehlike, M., Castillo, C.: Reducing disparate exposure in ranking: {A} learning
  to rank approach.
\newblock In: Y.~Huang, I.~King, T.~Liu, M.~van Steen (eds.) {WWW} '20: The Web
  Conference 2020, Taipei, Taiwan, April 20-24, 2020, pp. 2849--2855. {ACM} /
  {IW3C2} (2020).
\newblock \doi{10.1145/3366424.3380048}.
\newblock \urlprefix\url{https://doi.org/10.1145/3366424.3380048}

\bibitem{DBLP:journals/datamine/Zliobaite17}
Zliobaite, I.: Measuring discrimination in algorithmic decision making.
\newblock Data Min. Knowl. Discov. \textbf{31}(4), 1060--1089 (2017).
\newblock \doi{10.1007/s10618-017-0506-1}.
\newblock \urlprefix\url{https://doi.org/10.1007/s10618-017-0506-1}

\end{thebibliography}

\vspace{5mm}

\noindent {\bf Ludovico Boratto} is Senior Research Scientist in the Data Science and Big Data Analytics research group at Eurecat (Spain). His research interests focus on recommender systems and on their impact on the different stakeholders, both considering accuracy and beyond-accuracy evaluation metrics. The results of his research have been published in top-tier conferences and journals. His research activity also brought him to give talks and tutorials at top-tier conferences (e.g., ACM RecSys 2016, IEEE ICDM 2017) and research centers (Yahoo! Research). He is editor of the book ``Group Recommender Systems: An Introduction'', published by Springer. He is editorial board member of the ``Information Processing \& Management'' (Elsevier) and ``Intelligent Information Systems'' (Springer) journals, and guest editor of several journal’s special issues. He is regularly part of the program committee of the main Data Mining and Web conferences, such as RecSys, KDD, SIGIR, WSDM, ICWSM, and TheWebConf. In 2012, he got a Ph.D. at the University of Cagliari (Italy), where he was research assistant until May 2016. In 2010 and 2014 he spent 10 months at Yahoo! Research in Barcelona as a visiting researcher. He is member of ACM and IEEE.

\vspace{2mm}

\noindent {\bf Gianni Fenu} is Full Professor of Computer Science at the Department of Mathematics and Computer Science of the University of Cagliari (Italy). He is the Vice-Rector of the University of Cagliari and also the Director of the E-Learning for Didactic Innovation Center. He received the Laurea Degree in Engineering from the University of Cagliari in 1985, and he joined the University of Cagliari in 1988. He teaches courses of Computer Networks in the Bachelor's Degree in Computer Science, and Digital Transformation courses in the Master's Degree in Computer Science. His research interests include recommender systems, complex networks, e-learning, and biometrics. He is author and co-author of more than 130 papers published in scientific journals or proceedings of refereed conferences. Many research projects were managed by him as scientific responsible.

\vspace{2mm}

\noindent {\bf Mirko Marras} is Postdoctoral Researcher at the Digital Vocational Education and Training - Machine Learning for Education Laboratory of EPFL (Switzerland). His current research focuses on personalization, machine learning, and education, with special attention to beyond-accuracy aspects. He has co-authored papers in top-tier international journals, such as Information Processing \& Management (Elsevier), Computers in Human Behavior (Elsevier), and IEEE Cloud Computing. His research activity also brought him to give talks (e.g., ECIR 2019, EEE 2020), demos (e.g., TheWebConf 2018, INTERSPEECH 2019), and tutorials (e.g., ICDM 2020, WSDM 2021) at several international conferences. He is part of the program committee of top-tier international conferences, such as ACL, AIED, ECML-PKDD, EDM, EMNLP, ITICSE, RecSys, SIGIR, and UMAP. He has co-chaired the BIAS workshop at ECIR 2020 and 2021. He has served as a guest editor for special issues of the Information Processing \& Management and Future Generation Computer Systems journals (Elsevier). He has been involved in several national and European projects (e.g., iLearnTV, DECODE, EDUC). He is a member of several (inter)national associations, including AIxIA, IEEE, and ACM.

\end{document}